\documentclass[journal]{IEEEtran}
\newcommand{\tool}{\textsc{Casandra}}
\newcommand{\soa}{state-of-the-art}
\usepackage{amsmath,amssymb}
\usepackage{multirow}
\usepackage[table,xcdraw]{xcolor}
\usepackage[utf8]{inputenc}
\usepackage{cleveref}
\crefname{section}{§}{§§}
\Crefname{section}{§}{§§}
\usepackage{float}
\usepackage{graphicx}
\usepackage{hyperref}
\usepackage{color}
\hypersetup{
	colorlinks   = true,
	citecolor    = blue,
	urlcolor     = blue
}

\makeatletter
\DeclareUrlCommand\ULurl@@{%
	\def{\scriptsize ($ \pm 0.00 $)}Font{\ttfamily\color{blue}}%
	\def{\scriptsize ($ \pm 0.00 $)}Left{\uline\bgroup}%
	\def{\scriptsize ($ \pm 0.00 $)}Right{\egroup}}
\def\ULurl@#1{\hyper@linkurl{\ULurl@@{#1}}{#1}}
\DeclareRobustCommand*\ULurl{\hyper@normalise\ULurl@}
\makeatother

\usepackage{array}
\newcolumntype{L}[1]{>{\raggedright\let\newline\\\arraybackslash\hspace{0pt}}m{#1}}
\newcolumntype{C}[1]{>{\centering\let\newline\\\arraybackslash\hspace{0pt}}m{#1}}
\newcolumntype{R}[1]{>{\raggedleft\let\newline\\\arraybackslash\hspace{0pt}}m{#1}}

\usepackage{caption}
\usepackage{subcaption}
\usepackage{amsfonts}
\usepackage{fixltx2e}
\usepackage{bm}
\usepackage{amsmath}
\usepackage{graphicx}
\usepackage{subcaption}
\usepackage{amssymb}
\usepackage{tikz}
\usepackage{array}
\usepackage{xcolor}
\usepackage[mathscr]{euscript}
\usepackage{mathrsfs}

\usetikzlibrary{shapes.geometric, arrows}
\tikzstyle{ADG} = [ellipse, minimum width=1cm, minimum height=.75cm,text centered, draw=black, fill=pink!30]
\tikzstyle{PDG} = [ellipse, minimum width=1cm, minimum height=.75cm,text centered, draw=black, fill=yellow!20]
\tikzstyle{SSCFP} = [ellipse, minimum width=1cm, minimum height=.75cm,text centered, draw=black, fill=green!10]
\tikzstyle{Ins} = [ellipse, minimum width=1cm, minimum height=.75cm,text centered, draw=black, fill=brown!10]
\tikzstyle{Signs} = [ellipse, minimum width=1cm, minimum height=.75cm,text centered, draw=black, fill=orange!10]

\usepackage{listings}
\lstdefinestyle{customc}{
	belowcaptionskip=1\baselineskip,
	breaklines=true,
	frame=L,
	xleftmargin=\parindent,
	language=Java,
	showstringspaces=false,
	basicstyle=\footnotesize\ttfamily,
	keywordstyle=\bfseries\color{green!40!black},
	commentstyle=\itshape\color{purple!40!black},
	identifierstyle=\color{blue},
	stringstyle=\color{orange},
}

\usepackage{pifont}
\usepackage {algorithm,algorithmicx}
\usepackage{algpseudocode}

\usepackage{verbatim}

\usepackage{enumitem}
\setlist[enumerate]{leftmargin=*}
\usepackage[space]{cite}
\usepackage{footmisc}
\usepackage{footnote}
\usepackage[nodisplayskipstretch]{setspace}
\usepackage[normalem]{ulem}

\makeatletter
\def\therule{\makebox[\algorithmicindent][l]{\hspace*{.5em}\vrule height .75\baselineskip depth .25\baselineskip}}%

\newtoks\therules% Contains rules
\therules={}% Start with empty token list
\def\appendto#1#2{\expandafter#1\expandafter{\the#1#2}}% Append to token list
\def\gobblefirst#1{% Remove (first) from token list
	#1\expandafter\expandafter\expandafter{\expandafter\@gobble\the#1}}%
\def\LState{\State\unskip\the\therules}% New line-state
\def\pushindent{\appendto\therules\therule}%
\def\popindent{\gobblefirst\therules}%
\def\printindent{\unskip\the\therules}%
\def\printandpush{\printindent\pushindent}%
\def\popandprint{\popindent\printindent}%

\algdef{SE}[WHILE]{While}{EndWhile}[1]
{\printandpush\algorithmicwhile\ #1\ \algorithmicdo}
{\popandprint\algorithmicend\ \algorithmicwhile}%
\algdef{SE}[FOR]{For}{EndFor}[1]
{\printandpush\algorithmicfor\ #1\ \algorithmicdo}
{\popandprint\algorithmicend\ \algorithmicfor}%
\algdef{S}[FOR]{ForAll}[1]
{\printindent\algorithmicforall\ #1\ \algorithmicdo}%
\algdef{SE}[LOOP]{Loop}{EndLoop}
{\printandpush\algorithmicloop}
{\popandprint\algorithmicend\ \algorithmicloop}%
\algdef{SE}[REPEAT]{Repeat}{Until}
{\printandpush\algorithmicrepeat}[1]
{\popandprint\algorithmicuntil\ #1}%
\algdef{SE}[IF]{If}{EndIf}[1]
{\printandpush\algorithmicif\ #1\ \algorithmicthen}
{\popandprint\algorithmicend\ \algorithmicif}%
\algdef{C}[IF]{IF}{ElsIf}[1]
{\popandprint\pushindent\algorithmicelse\ \algorithmicif\ #1\ \algorithmicthen}%
\algdef{Ce}[ELSE]{IF}{Else}{EndIf}
{\popandprint\pushindent\algorithmicelse}%
\algdef{SE}[PROCEDURE]{Procedure}{EndProcedure}[2]
{\printandpush\algorithmicprocedure\ \textproc{#1}\ifthenelse{\equal{#2}{}}{}{(#2)}}%
{\popandprint\algorithmicend\ \algorithmicprocedure}%
\algdef{SE}[FUNCTION]{Function}{EndFunction}[2]
{\printandpush\algorithmicfunction\ \textproc{#1}\ifthenelse{\equal{#2}{}}{}{(#2)}}%
{\popandprint\algorithmicend\ \algorithmicfunction}%

\makeatother

\begin{document}
	\title{Context-aware, Adaptive and Scalable Android Malware Detection through Online Learning (extended version)}
	
	\author{Annamalai~Narayanan,
		Mahinthan~Chandramohan,
		Lihui~Chen,
		and~Yang~Liu
		\\ Nanyang Technological University, Singapore.
		\\annamala002@e.ntu.edu.sg, \{mahinthan, elhchen, yangliu\}@ntu.edu.sg}

	\markboth{Journal of \LaTeX\ Class Files,~Vol.~11, No.~4, December~2012}%
	{Shell \MakeLowercase{\textit{et al.}}: Bare Demo of IEEEtran.cls for Journals}
	
	\maketitle
	
	\begin{abstract}
		It is well-known that Android malware constantly evolves so as to evade detection. This causes the entire malware population to be non-stationary. Contrary to this fact, most of the prior works on Machine Learning based Android malware detection have assumed that the distribution of the observed malware characteristics (i.e., features) does not change over time. In this work, we address the problem of \textit{malware population drift} and propose a novel online learning based framework to detect malware, named \tool{} (\uline{C}ontext-aware, \uline{A}daptive and \uline{S}calable \uline{ANDR}oid m\uline{A}lware detector). In order to perform accurate detection, a novel graph kernel that facilitates capturing apps’ security-sensitive behaviors along with their context information from dependency graphs is proposed. Besides being accurate and scalable, \tool{} has specific advantages: (i) being adaptive to the evolution in malware features over time (ii) explaining the significant features that led to an app’s classification as being malicious or benign. In a large-scale comparative analysis, \tool{} outperforms two \soa{} techniques on a benchmark dataset achieving 99.23\% F-measure. When evaluated with more than 87,000 apps collected in-the-wild, \tool{} achieves 89.92\% accuracy, outperforming existing techniques by more than 25\% in their typical batch learning setting and more than 7\% when they are continuously retained, while maintaining comparable efficiency. %Besides this, \tool{} demonstrates excellent ability to adapt to the drift in real-world malware over time and great potential as a reliable framework to perform market-scale analysis. 
	\end{abstract}
	
	
	\IEEEpeerreviewmaketitle

	\section{Introduction}
	\label {sec:intro}
	\IEEEPARstart{I}{n} recent times, malware detection for mobile platforms such as Android has evolved as one of the challenging problems in the field of cyber-security. The number of new Android malware applications (apps) have grown tremendously in recent years. For instance, Symantec reports \cite{report} discovering 430 million new malware in 2015 which is a 36\% increase over 2014. Also their capabilities have grown from simple phone cloning, sending premium-rated SMS to complex botnets, cryptolocker and ransomware \cite{droidscribe,droidseive,prescience,androidsurvey}. Besides this, attackers continuously enhance the sophistication of malware to evade novel detection techniques. The sheer volume, growth rate and evolution of sophisticated Android malware highlights an imperative need for developing sound and scalable automated malware detection techniques \cite{prescience,Drebin,droidscribe,androidsurvey,CSBD}.
	
	\textbf{Machine Learning based malware detection.}
	For over a decade, Machine Learning (ML) techniques have been predominantly used to perform malware detection in various platforms (such as Windows and Android) \cite{Drebin,droidscribe,droidseive,CSBD,Adagio,DroidMiner,AppContext,DroidSift,MLMalDetect,ConDrift,Mudflow,androidsurvey,prescience}. This is because, ML methods automatically learn the characteristics that distinguish malicious behavior, when trained using a collection of malware and benign samples making them amenable for automated detection. ML based approaches extract semantic features from apps’ behaviors and apply standard classification algorithms (e.g., Support Vector Machine (SVM), Random Forests (RFs), etc.) to perform binary classification. These approaches typically use features such as system calls/Application Programming Interfaces (APIs) invoked, resources and privileges used, control- and data-flows inside apps’ execution to detect malicious behavior patterns \cite{DroidMiner,DroidSift,Adagio,Mudflow,CSBD,MLMalDetect}. These semantic features are extracted through static \cite{Mudflow,Drebin,CSBD,Adagio,AppContext,DroidSift} and dynamic \cite{droidscribe,Crowd} program analysis.

	\subsection{Malware detection using graph representations}
	\label{ss:intro_graph}

	\textbf{Malware variants.}
	A major reason for the tremendous growth rate in malware is the production of malware variants. Typically, the attackers produce large number of variants of the same malware by resorting to techniques such as variable renaming and junk code insertion \cite{androidsurvey,s&p,Mudflow}. These variants perform same malicious functionality, with apparently different syntax, thus evading syntax-based detectors. However, higher level semantic representations such as call graphs, control- and data-flow graphs, control-, data- and program-dependency graphs mostly stay similar even when the code is considerably altered \cite{DroidMiner,Adagio,AppContext,CSBD,DroidSift}. In this work, we use a common term, Program Representation Graph (PRG) to refer to any of the aforementioned graphs.
	As PRGs are resilient against variants, many works in the past have used them to perform malware detection.
	%As they are resilient against variants, many works in the past have extracted several semantic features from PRGs and subsequently used them to perform malware detection. 
	In essence, such works cast malware detection as a graph classification problem and apply existing graph mining and classification techniques \cite{Adagio,AppContext,DroidSift,s&p}. Some methods such as \cite{s&p,Adagio,MLMalDetect} note that ML classifiers are readily applicable on data represented as vectors and attempt to encode PRGs as feature vectors. 
	
	\textbf{Graph Mining and Kernels.} Many existing works such as \cite{acts} and \cite{camas} use off-the-shelf graph mining algorithms on PRGs for malware detection. However, it has to be noted that the scale of malware detection problem is such that we have millions of samples already and thousands streaming in every day. 
	Many classic graph mining based approaches (e.g., \cite{s&p}) are NP hard and have severe scalability issues, making them impractical for real-world malware detection. %In fact, both the aforementioned approaches have been evaluated with a dataset comprising less than 1300 malware samples.
	
	%\textbf{Graph Kernels.}
	On the other hand, one of the increasingly popular approaches in ML for graph-structured data is the use of graph kernels. 
	These graph kernels could be used together with a kernel classifier (e.g., SVM) to perform graph classification \cite{WLK}.
	Recently, efficient and expressive graph kernels such as \cite{WLK}, \cite{NHGK} and \cite{NSPDK} have been proposed and widely adopted in many application domains (e.g, computer vision \cite{CVGK}, chemoinformatics \cite{Molec}, etc.). These kernels have been known to operate in linear-time and have produced accurate results in many real-world applications. Therefore, it just suffices to apply any of these kernels on suitable PRGs and we have an effective, scalable and ready-to-use malware detector. 
	%Recently, two approaches \cite{Adagio} and \cite{MLMalDetect} have successfully demonstrated using these general purpose graph kernels for Android malware detection. 
	However, as we explain later, these general purpose graph kernels do not take many problem-specific constraints and hence yield suboptimal accuracies.
	
	\subsection{Challenges in ML based malware detection}
	\label{ss:intro_chall}
	Almost all ML based Android malware detection techniques (incl. the aforementioned ones) operate in a \textit{batch-learning} setting and use off-the-shelf batch learners like SVM or RFs. Meaning, the detection model is built using a batch of labeled benign and malware samples and is subsequently used to predict whether a given new sample is benign or malicious. 
	
	In general, these ML based approaches are typically plagued by four challenges that make them unsuitable for large-scale real-world malware detection: 
	
	\textbf{(C1) Population drift.} 
	Though batch-learning based solutions are promising, their success is predicated on an important assumption that may not hold for the malware detection problem. Meaning, \textit{they assume that the malware population (i.e., training data) used to build the detection model does not change over time}. However, malware does not fit this profile. The entire population of malware is constantly evolving due to various reasons such as exploiting new vulnerabilities and evading novel detection techniques. This evolution has a profound impact on malware characteristics and thereby on the features used by these ML models. This makes the collection of malware identified today unrepresentative of the ones generated in the future. This phenomenon is an epitome of \textit{population drift} \cite{ConDrift,prescience}.
	%This population drift leads to a concept drift that needs to handled by the ML models \cite{ConDrift,prescience}. 
	
	\textbf{(C2) Volume.} 
	Since the malware population grows at an alarming rate, a scalable classifier is of paramount importance for real-world malware detection. In order to keep abreast with drifting population, batch learners have to be frequently retrained with huge volumes of data. Hence they pose severe scalability issues when used in the Android malware detection context where we have thousands of apps streaming in every day. Retraining frequently with such a volume renders them computationally impractical.

	\textbf{(C3) Explainability.} In general, these ML based solutions just predict the labels of a given sample without offering insights or explanations into how those predictions are arrived at. In other words, they act as \textit{black-box} solutions. However, for malware detection models, understanding the reasons behind their predictions is important in assessing their trustworthiness. This is fundamental if one plans to take action such as deploying a new model or studying malware evolution  based on these predictions.
	
	\textbf{(C4) Expressiveness.} 
	%Only the approaches that leverage on PRGs face this challenge.
	PRGs are known to be complex and expressive data structures that characterize topological relationships among program entities. Representing them as vectors or other formats amenable for applying ML algorithms is a non-trivial task \cite{s&p}. 
	In many cases such representations fail to capture all the vital information from PRGs, thus loosing their expressiveness.
	For instance, solutions like \cite{DroidMiner}, \cite{Adagio} and \cite{massvet} capture the topological neighborhood (i.e., structural) information from PRGs and detect security-sensitive behaviors through analyzing them. 
	However, as revealed by recent works such as \cite{DroidSift} and \cite{AppContext}, an important contextual factor that distinguishes malice is whether or not the user is aware of such behaviors. 
	%In other words, whether such sensitive operations happen in the \textit{user-aware} or \textit{user-unaware} context.
	Unfortunately, the above-mentioned methods which capture structural information well, fail to capture the contextual information and this leads to raising false alarms even when sensitive operations are performed with users' consent. The general purpose graph kernels such as \cite{rw,frw,sp,graphlet,WLK,NHGK,NSPDK} also suffer with the same drawback.
	%On the other hand, solutions like AppContext \cite{AppContext} capture contextual information from individual nodes without their topological neighbourhood information. 
	%With such loss of expressiveness, attacks that span across multiple nodes could not be effectively detected. 
	%As mentioned earlier, some of these issues could be solved by deploying an expressive graph kernels. However, a major problem in using these general purpose graph kernels on PRGs is that, they are not designed to take domain-specific observations into account. For instance, many existing graph kernels such as \cite{WLK}, \cite{NSPDK} and \cite{dgk} can capture and compare structural information from PRGs effectively. However, these kernelsare not designed to capture the contextual information, as it is a strong application- and domain-specific requirement and hence fail to do so.
	
	In summary, challenges C1, C2 and C3 impair all ML based approaches and C4 deters graph based learning approaches.
	
	%\textbf{Online Learning} Online learning can fix C1, C2. Linear online models can fix C3. Novel graph kernel is needed for C4.

	%Research Gap.  To address this, we develop a novel graph kernel which is capable of capturing both the aforementioned types of information. For similar domain-specific reasons, researchers from other fields such as computer vision [], bio- and chemoinformatics [] have developed a number of kernels that specifically suit their applications. Despite graphs being natural representations of programs and amenable for various activities, the program analysis research community has not devoted significant attention to development of domain-specific graph kernels. We take the first step towards this, by developing a kernel on PRGs which specifically suits our task of malware detection.

	\subsection{Our Approach}
	\label{ss:intro_approach}
	We take these four challenges into consideration and propose \tool (\uline{C}ontext-aware, \uline{A}daptive and \uline{S}calable \uline{ANDR}oid m\uline{A}lware detection) framework based on online learning, where we continuously retrain the model upon receiving each labeled sample and make prediction of a new sample using the updated model. 
	%We demonstrate that online learning based solutions are better suited for practical large-scale automated malware detection for two reasons: (1) the detection model needs to adapt to changes in malware features over time, automatically and (2) online learning based solution can process large numbers of apps more efficiently than batch methods
	\tool{} is developed with the four following design goals:\\
	\textbf{1. Accuracy.} Accuracy of \tool{}, which is a PRG based approach depends on how well it retains PRG's expressiveness. In other words, it depends on how well the approach addresses challenge C4. To this end, we leverage on our previous work \cite{cwlk} and use the Contextual Weisfeiler-Lehman kernel (CWLK) that is specifically designed to capture both structural and contextual information from PRGs.\\
	% from   develop a novel graph kernel which is capable of capturing both structural and contextual information from PRGs. 
	%More specifically, we propose a method to enrich the feature space of a graph kernel that inherently captures structural information with contextual information. We apply this feature enrichment idea on a state-of-the-art graph kernel, namely, Weisfeiler-Lehman kernel (WLK) \cite{WLK} to obtain the Contextual Weisfeiler-Lehman kernel (CWLK). CWLK associates to each sub-structure feature of WLK a piece of information about the context under which the sub-structure is reachable in the course of execution of the program. A sub-structure appearing in two different PRGs will match only if it is reachable under the same context in both PRGs. We show that for the malware detection problem, CWLK is more expressive and hence more accurate than WLK while maintaining comparable efficiency.% (see \S \ref{ss:rq1}). 
	%This addresses challenge C4.
	\textbf{2. Efficiency.} \tool{} achieves its efficiency through the combined use of a scalable graph kernel (i.e., CWLK) 
	and a \soa{} online classifier, namely, Confidence Weighted (CW) algorithm \cite{cw}. %As we report through our experimental results (see \S \ref{ss:rq2}), \tool's efficiency is comparable to that of \soa{} solutions.
	This addresses challenge C2.\\
	\textbf{3. Adaptiveness.} \tool{} automatically adapts to malware population drift through its use of online classifier and thus addressing challenge C1.\\ %As we report through our experimental results (see \S \ref{ss:rq4} and \S \ref{ss:rq1}), this adaptiveness helps \tool{} to achieve a 25\% improvement in accuracy over \soa{} solutions.
	\textbf{4. Explainability.} Since \tool{} uses a linear classifier along with CWLK which permits explicit feature vector representation of PRGs, we are able to categorically identify the PRG subgraph features which contribute to its predictions. This makes \tool's predictions explainable thus addressing challenge C3. %As we report through our experimental results (see \S \ref{ss:rq3}), we show that \tool's explanations are better and reflect the malice behavior more closely than the \soa{} solutions.

	\textbf{Experiments.} \tool{} is evaluated through large-scale experiments both on benchmark and a recent real-world dataset of more than 87,000 apps. 
	%With the proposed CWLK, we extract more than 2 million features from these apps and perform explainable malware detection. 
	%\tool's accuracy and efficiency are compared against those of state-of-the-art malware detection techniques. 
	\tool{} achieves 89.92\% accuracy on the real-world dataset, outperforming two state-of-the-art techniques by more than 25\% in their typical batch-learning setting and more than 7\% when they are continuously retrained, while maintaining comparable efficiency.
	% (see \S \ref{ss:rq1} for details). 
	%Also, \tool's efficiency is comparable to that of state-of-the-art techniques.
	% (see \S \ref{ss:rq2} for details). 
	%Through its explainable detection process, \tool{} offers insights into its predictions.
	%also study Android malware evolution how their features get complex and sophisticated over time.
	% (see \S \ref{ss:rq3} for details). 
	%Finally, we demonstrate \tool's adaptiveness and how it handles malware population drift. %We show that continuous retraining over newly emerging features is crucial for adapting the detection model to detect new or evolving malware.
	% (see \S \ref{ss:rq4} for details).
	Through further experiments, we demonstrate \tool's explainable detection process and its adaptiveness. 
	
	\textbf{Contributions.} 
	On one hand, in our recent work \cite{cwlk}, we developed CWLK, a graph kernel that is specifically designed to perform malware detection using PRGs\footnote{To the best of our knowledge, CWLK was the first graph kernel specifically addressing a problem from the field of program analysis.}. CWLK captures both contextual and structural information, enabling it to achieve high accuracy in a batch learning setting. On the other hand, another recent work of ours, DroidOL \cite{droidol}, demonstrated that online learning based solutions are better suited for large-scale real-world automated malware detection than batch learning methods. However, DroidOL used a general purpose kernel which can only capture structural information from PRGs. In this work, we bring together the best of both the worlds. More specifically, \textit{we combine a kernel which is specifically designed to cater the malware detection task (i.e., CWLK), with a learning paradigm that suits the task extremely well (i.e.,online learning)}. 
	To the best of our knowledge \tool{} is the first framework that leverages on both task-specific  kernel and online learning to perform malware detection. Besides this, we have made the three following new contributions in \tool{}:
	\begin{itemize}[leftmargin=*]
		\setlength\itemsep{0em}
		\item Performing explainable malware detection is an unique feature of \tool. As we demonstrate through our experiments, \tool's explanations are more comprehensive and semantically closer to the malicious behaviors than those of \soa{} approaches. Both \cite{cwlk} and \cite{droidol} were not demonstrated to possess this.
		
		\item Adaptiveness is another important trait of \tool{} which is not prevalent in any existing approach. We have designed new experiments to thoroughly demonstrate how \tool{} adapts to malware population drift (see \S \ref{ss:rq4} for details). Though DroidOL \cite{droidol} possessed this quality, it was neither experimentally verified nor demonstrated.
		
		\item We also replaced the online learner used in DroidOL \cite{droidol} with a more recent \soa{} online learner. This resulted in significantly better accuracies\footnote{DroidOL outperformed \soa{} batch-learning approaches by nearly 20\% (without retraining) and 3\% (with retraining), whereas, \tool{} outperform the same by more than 25\% (without retraining) and 7\% (with retraining) accuracies, respectively. We believe, this is due to the changes in the choice of the kernel and the classifier.}.
	\end{itemize}

	%In summary, this paper’s contributions are as follows:
	%\begin{itemize}
	%\item We propose and develop \tool, an accurate, scalable, adaptive and explainable Android malware detection framework based on online learning, where we do not assume that the malware population is stationary. %To the best of our knowledge, we are the first to propose an online learning framework and demonstrate its capabilities to handle population drift while performing malware detection.
	%
	%%\item We develop a graph kernel that captures both structural and contextual information from PRGs to perform accurate and scalable malware detection. To best of our knowledge, this is the first graph kernel specifically addressing a problem from the field of program analysis.
	%
	%\item We interpret and explain the detection results of the apps in our dataset in terms of malicious attacks performed and thus study the evolution of Android malware over time.
	%
	%\item Through large-scale experiments and comparative analysis on both benchmark and real-world datasets, we show that \tool{} outperforms state-of-the-art malware detection solutions in terms of accuracy, while maintaining high efficiency.
	%
	%
	%%\item We make an efficient implementation of the proposed framework (along with the dataset information) publicly available\footnote{\url{https://github.com/MLDroid}}.
	%
	%\end{itemize}
	
	\textbf{Organization.} The paper is organized as follows: We begin by introducing the background and motivations for our framework's design in \S\ref{sec:bgps}. The proposed \tool{} framework is presented in \S\ref{sec:method}.
	%(details on the proposed kernel and online learning using CW are presented subsections \S\ref{ss:cwlk} and \S\ref{ss:ol}, respectively). 
	The experimental design and implementation details are furnished in \S \ref{sec:ed}. \tool's evaluation results and discussions are presented in section \S\ref{sec:r_d}. Related Work, Limitations and Conclusions are provided in \S \ref{sec:rw}, \S\ref{sec:lim} and \S\ref{sec:conc}, respectively.
	
	\begin{figure*}[t]
		\centering
		\includegraphics[height=3cm,width=18cm]{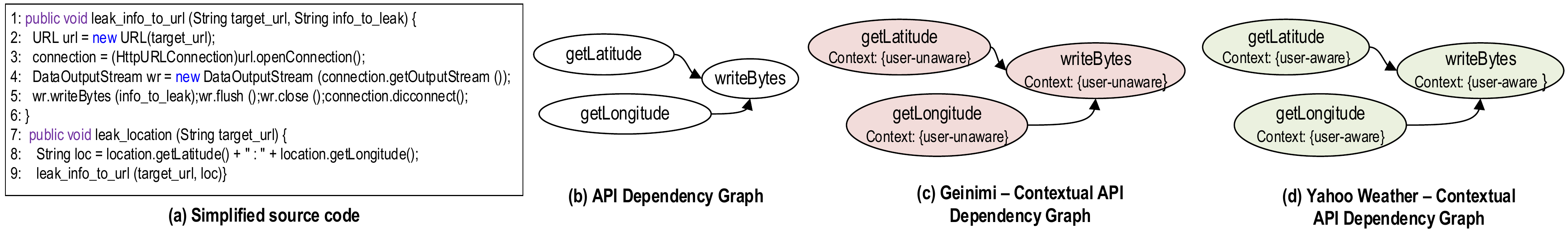}
		\caption{\small Location information being leaked in \textit{Geinimi} (malware) and \textit{Yahoo Weather} (benign) apps. (a) code snippet corresponding to leaking location information in both the apps. (b) ADG corresponding to the location leak. (c) \textit{Geinimi's} CADG illustrating that it leaks information without the user's knowledge. (d) \textit{Yahoo Weather's} CADG illustrating that it leaks information with the user's knowledge. \label {fig:why-cwlk}}
	\end{figure*}

	\section{Background and Motivation}
	\label{sec:bgps}
	
	In this section, the background on Android malware detection and motivations for the two main components of the \tool{} framework, namely, CWLK and online learning are presented. Specifically, we discuss the two following motivations: (1) why considering structural information alone from PRGs is insufficient to determine the maliciousness of a sample and how supplementing it with contextual information helps to increase the detection accuracy and (2) why using batch learning is impractical for building a real-world malware detector and how the use of online learning alleviates such practicality issues. For the preliminaries on kernel methods and graph kernels interested readers are referred to Appendix \ref{app:kmgk}.  
	
	\subsection{Motivations for CWLK}
	\label{ss:mot-cwlk}
	
	To demonstrate the necessity of CWLK, we use a real-world Android malware from the \textit{Geinimi} family which steals users’ private information and contrast its behavior with that of a well-known benign app, \textit{Yahoo Weather}.  
	
	\textbf{Geinimi’s execution.} The app is launched through a background event such as receiving a SMS or call. Once launched, it reads the user’s personal information such as geographic location and contacts and leaks the same to a remote server. The (simplified) malicious code portion pertaining to the location information leak is shown in Fig. \ref{fig:why-cwlk} (a). The method {\tt leak\_location} reads the geographic location through {\tt getLatitude} and {\tt getLongitude} APIs. Subsequently, it calls {\tt leak\_info\_to\_url} method to leak the location details (through {\tt DataOutputStream.writeBytes}) to a specific server. The API dependency graph (ADG)\footnote{The detailed procedure for constructing the ADG is provided later in \S \ref{ss:sa}.} corresponding to the code snippet is shown in Fig. \ref{fig:why-cwlk} (b). The nodes in ADG are labeled with the sensitive APIs that they invoke and the edges denote the control-flow dependencies. 
	
	\textbf{Yahoo Weather’s execution.} On the other hand, \textit{Yahoo Weather} could be launched only by user’s interaction with the device (e.g., by clicking the app’s icon on the dash board). The app then reads the user’s location and sends the same to its weather server to retrieve location-specific weather predictions. Hence, ADG portions of \textit{Yahoo Weather} is same as that of \textit{Geinimi}.
	
	\textbf{Contextual information.} From the explanations above, it is clear that both the apps leak the same information in the same fashion. However, what makes \textit{Geinimi} malicious is the fact that its leak happens without the user’s consent. In other words, unlike \textit{Yahoo Weather}, \textit{Geinimi} leaks private information through an event which is not triggered by user’s interaction. We refer to this as a leak happening in \textit{user-unaware} context. On the same lines, we refer to \textit{Yahoo Weather’s} leak as happening in \textit{user-aware} context. 
	As explained in \cite{DroidSift} and \cite{AppContext}, in the case of Android apps, one could determine whether a PRG node is reachable under \textit{user-aware} or \textit{user-unaware} context by examining its entry point nodes.
	%As explained in \cite{AppContext} and \cite{DroidSift}, in the case of Android apps, one could determine reachability context of a PRG node by examining its entry point nodes.
	Following this procedure we add the context as an attribute to every ADG node. This context annotated ADG of \textit{Geinimi} and \textit{Yahoo Weather} are shown in Fig. \ref{fig:why-cwlk} (c) and (d), respectively.

	\textbf{Requirements for effective detection.} From the aforementioned example the two key requirements that make a malware detection process effective can be identified: 
	
	\textbf{(R1) Capturing structural information.} Since malicious behaviors often span across multiple nodes in PRGs, just considering individual nodes (and their attributes) in isolation is not enough. For instance, in the case of \textit{Geinimi}, the privacy leak attack spans across three ADG nodes. Therefore, capturing the structural (i.e., neighborhood) information from PRGs is of paramount importance. 
	
	\textbf{(R2) Capturing contextual information.} Considering just the structural information without the context is not enough to determine whether a sensitive behavior is triggered with or without users' knowledge. For instance, if structural information alone is considered, the features of both \textit{Geinimi} and \textit{Yahoo Weather} apps become identical, thus making the latter a false positive. Hence, it is important for the detection process to capture the contextual information as well to make it more accurate.
	
	Many existing graph kernels could address the first requirement well. However, the second requirement which is more domain-specific makes the problem particularly challenging. To the best of our knowledge, none of the existing graph kernels support capturing this reachability context information. 
	In summary, this gives us a clear motivation to develop a new kernel that specifically addresses our two-fold requirement.
	% which facilitates explicit feature vector representations of PRGs.

	\subsection{Motivations for using Online Learning}
	\label{ss:mot-ol}

	As foreshadowed in \S \ref{sec:intro}, the three main motivations for using online learning instead of batch learning in \tool{} are:\\
	% (1) handling population drift in Android malware \\
	% (2) handling large volumes of streaming benign and malware samples represented as high dimensional data \\
	% (3) performing detection on data that streams-in at real-time.
	(1) handling population drift, (2) handling large volumes of high dimensional data, and (3) performing detection on data that streams-in at real-time.
	Malware detection-specific justifications on how online learning helps to address these challenges are presented below.
	\begin{figure*}[t]
		\centering
		\includegraphics[height=3cm,width=16cm]{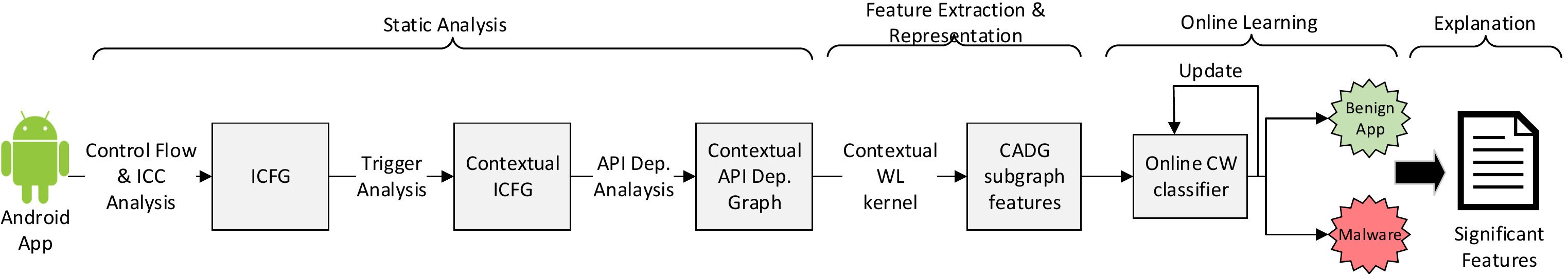}
		\caption{\small \tool: Framework overview \label {fig:framework}}
	\end{figure*}
	
	\textbf{Handling population drift.}
	As reported in \cite{Genome}, the attack and the evasion strategies of malware have evolved significantly, ever since the inception release of Android.{\color{black}{ Also, one could observe the following pattern in this evolution: \textit{attacks that were popular at some point in time could not sustain forever}.}} For instance, the \textit{BaseBridge} family of malware leveraged on two root exploits, namely, \textit{RATC} and \textit{Zimperlich} to escalate its privileges \cite{Genome}. Once the corresponding vulnerabilities were patched and the AV vendors became capable enough to detect such attacks, \textit{BaseBridge's} propagation and sustenance were affected, ultimately resulting in its extinction. This leads to the following observation: \textit{several malware families emerge, flourish and fade-away over time due to various domain-specific reasons}. 
	
	From an ML based malware detection viewpoint, this leads to emergence, dominance and disappearance of semantic features that characterize these attacks and evasion strategies. This results in the three following phenomena that happen over time: 
	(1) new features emerge %over time %(e.g., a new attack being introduced leveraging on a newly reported vulnerability/newly introduced APIs leads to the emergence of new subgraph patterns corresponding to those attacks) 
	(2) the significance of features vary %over time %(e.g., features characterizing attacks that fade-away due to above-mentioned reasons will loose their importance) and 
	(3) the cumulative number of features keeps monotonically increasing.
	This causes the underlying distribution of malware samples to change over time leading to what is known as a \textit{concept drift}. As a result of this drift, the predictions of a static model might become less accurate in due course of time. Therefore, detectors need to adapt to such changes quickly and accurately.

	A straightforward solution to this problem is to detect the magnitude and direction of drift in the concept over time and retrain the batch models when the drift is significant \cite{prescience}. However, this approach would be hampered if: (1) the data arrives in large volumes at a rate too fast to detect, or (2) retraining the models at regular intervals is too expensive. Unfortunately, both these conditions are true in the real-world malware detection setting, as we review in detail below. {\color{black} On the other hand, online learners which are trained on a stream of samples as they arrive are naturally adaptive to evolving distributions mainly due to their learning mechanisms that continuously and efficiently update the model with the most recent examples.}
	
	\textbf{Handling large high-dimensional data.}
	Particularly in the malware detection problem, data is large not only in sample size, but also in feature size. For instance, on the large-scale dataset used in our evaluation, even a light-weight method such as \textsc{Drebin} \cite{Drebin} extracts and uses more than 1 million semantic features (see \S \ref{ss:why-ol}). In such big data contexts, the efficiency of a model in terms of both memory and time is of paramount importance in determining its practicality. 
	
	%In the case of such big data problems, the traditional batch learning algorithms fall short in low efficiency and poor scalability. More specifically, batch models pose two severe drawbacks that make them impractical in such circumstances, (1) high memory consumption and (2) large training durations.
	However, many batch learners (e.g., SVMs, RFs) typically take multiple passes over the training samples to optimize their parameters and yield the most accurate models within their hypothesis space. Often times, they are optimized to have long and computationally expensive training phases in order to perform  accurate and faster during testing. %This is because training is assumed to happen rarely, often only once. 
	Also, they might require large memory as they hold a batch 
	%(or a mini-batch) 
	of samples in memory facilitating them to take multiple passes. On the other hand, online learners take exactly one pass over training data. Meaning, they look at each sample that streams-in only once and learn from them, without storing or iterating over them. This makes the online learners highly efficient in terms of both memory and time. Therefore, online methods offer more practically tractable solutions for malware detection over their batch counterparts.

	\textbf{Handling real-time streaming.} Android has a vibrant and growing app ecosystem with a considerable number of third-party markets. Google Play \cite{Play}, the official market, host more than 2.4 million apps as of this writing. Meaning, nearly 1120 new apps are submitted to Google Play, on average every day. Also, popular third-party markets (e.g.,\cite{amazon-store}) receive a similar number of new submissions on a daily basis.% (e.g., Amazon market received 1665.8 apps on average every day in Sep 2016 \cite{amazon-store}). 
	
	Clearly, one important use-case of malware detection solutions like \tool{} is to deploy them to perform automated vetting of apps submitted to these markets. Therefore, such solutions must be scalable, always up-to-date in terms of their knowledge on attack and evasion strategies and ready for predicting and offering insights into apps as and when they stream in. In other words, these model could not be trained with an outdated set of samples and could not take outages or downtime to retrain themselves. Unlike online methods, the batch learners are susceptible to both these disadvantages, making them unsuitable for handling classification over a continuous, fast and voluminous stream of apps.

	\section{Methodology}
	\label{sec:method}
	
	The methodology of \tool{} framework designed to perform context-aware, scalable, adaptive and explainable malware detection is presented in this section. We begin by presenting the framework overview and subsequently, we explain each component of the framework.
	
	\subsection {\tool: Framework overview}
	\label{ss:ov}
	Fig. \ref{fig:framework} depicts \tool{} framework's overview.  The framework has four components as described below. \\
	\textbf{Static Analysis.} Our framework considers subgraphs from Contextual API Dependency Graphs (CADGs) as semantic features to perform malware detection. To this end, we first perform static program analysis to transform the given set of apps into their corresponding CADG representations. This procedure is explained in detail in section \S \ref{ss:sa}.\\
	\textbf{Feature Extraction \& Representation.} Once the CAGD of the apps in the dataset are constructed, those subgraphs which represent the security-sensitive events that happen in every app along with their context are extracted as using CWLK. We follow a Bag-of-Features (BoF) model \cite{WLK} to construct the feature vector of individual apps. The detailed procedure is explained in \S \ref{ss:cwlk}.\\
	\textbf{Online Learning.} Once the feature vectors of all the apps in the training set are built, we train a CW classifier with them to detect malware. CW classifier’s training and update procedures are explained in detail in \S \ref{ss:ol}. Subsequent to this training \tool{} is ready to perform malware detection at scale. It is important to note that the classifier is trained in an online fashion, meaning whenever a sample is presented, the classifier not only predicts its label but also updates the model based on the actual label of the sample. \\%In this sense, the classifier performs malware detection and simultaneously adapts to the changing trends in malware characteristics.\\
	\textbf{Explanation.} ML based malware detection solutions are usually \textit{black-box} methods as they do not explain why a particular sample is detected as malicious or benign. In \tool, we address this shortcoming as follows: besides detection \tool{} reports significant CADG subgraphs of an app that contribute to a detection. In most cases, these significant subgraphs reveal the app's characteristics related to malicious or benign behaviors. %Since \tool{} uses a linear classifier, we are able to determine the contribution of each feature to the final class prediction. 
	The detailed procedure is presented in \S \ref{ss:exp}.

	\subsection{Static Analysis}
	\label{ss:sa}
	CASANDRA considers CADG subgraphs as features which encompasses both structural and contextual information characterizing
	security sensitive operations from apps. We
	perform a comprehensive static analysis on a given app to
	construct its CADG. Our static analysis and CADG construction process is very similar to that of DroidSIFT \cite{DroidSift} and AppContext \cite{AppContext}. 
	
	The CADG construction workflow as depicted in Fig. \ref{fig:framework} involves three steps. Each of them are described below.
	
	\textbf{Step 1: Inter-procedural Control Flow Graph (ICFG) Construction.} The procedure mentioned in \cite{AppContext} is followed to construct the ICFG of a given app. Intuitively, the nodes of ICFG are the instructions in every method of the app. The ICFG edges are the intra- and inter-procedural control flows among these instructions. There two types of interprocedural invocations in Android apps: (1) method invocations and (2) Inter Component Communications (ICCs). Method invocations could be directly inferred through control flow analysis. However, ICCs are not straight forward as these invocations are facilitated by the underlying Android OS framework. In order to model such invocations, we leverage on the IC3\footnote{It is noted that both DroidSIFT and AppContext  used Epicc \cite{epicc} for their ICC analysis. We have replaced it with IC3 which is more recent and \soa.} tool \cite{IC3} and perform a precise ICC analysis.
	
	\textbf{Step 2: Contextual ICFG (CICFG) Construction.} During the course of execution of the program, the ICFG nodes could be reached by actions triggered by entities such as user or system. Trigger events are the external events such as users' interaction with app's user-interface (UI) and system changes such as receipt of SMS that trigger invocation of security-sensitive APIs. Trigger events connect security-sensitive behaviors to their "initiator" in the external environment (e.g., users or system). 
	DroidSIFT \cite{DroidSift} proposed a method to analyze the entry point nodes (i.e., PRG nodes that do not have any incoming edge) to identify and categorize the trigger events into UI and NON-UI triggers.
	We extend this procedure and categorize trigger events of ICFG nodes into the three following types:
	
	\begin{enumerate}
		\setlength\itemsep{0em}
		
		\item UI triggers: events triggered by interactions on apps’ UI (e.g,  Clicking UI buttons to make calls, etc.).
		
		\item NON-UI triggers: events initiated by the system state changes (e.g., receipt of SMS, change of signal strength), and events initiated by hardware related actions (e.g., pressing the HOME or BACK button).
		
		\item \textsc{Unresolved} triggers: Entry points of certain sensitive methods are not traceable by our static analysis. For instance, \textit{DroidKungFu}, a popular malware family has its malicious payload triggered by dynamically loaded code. However, similar to \cite{DroidSift} and \cite{AppContext}, our static analysis cannot model dynamic code loading and reflection  based triggers. %However, 
		%such triggers are only a minority and techniques such as
		%DroidSIFT \cite{DroidSift} chose to ignore them. 
		Therefore, we consider them as behaviors with \textsc{Unresolved} trigger events rather than ignoring them. 
	\end{enumerate}
	
	Understandably, each ICFG node is triggered by one or more of  these triggers. We consider ICFG nodes that are reachable through UI and NON-UI triggers to be in the \textit{user-aware} and \textit{user-unaware} contexts, respectively. Nodes that are reached through \textsc{Unresolved} triggers are considered to be in the '\textit{unresolved}' context.
	The reachability context(s) of every node is added as a node attribute. This transforms ICFG to CICFG. To formally define CICFG, we adopt and extend the definition of an ICFG from Harrold et al. \cite{ICFGPaper} as follows.
	
	\textbf{Definition 1 (CICFG).} $CICFG = (N_i, E_i, \xi)$ of an app $a$ is a directed graph in which each node $n \in N_i$ denotes an instruction in every method of $a$, and each edge $e(n_1, n_2) \in E_i$ denotes either an intra-procedural control-flow dependence from $n_1$ to $n_2$ or a calling relationship from $n_1$ to $n_2$. $\xi$ is a set of contexts though which every node $n \in N$ could be reached.
	
	\textbf{Step 3: CADG construction.}
	The CICFG captures the complete control-flow across every instruction in an app, along with context information. However, only a selected minority of these CICFG nodes will be security-sensitive (i.e., will invoke sensitive APIs). Therefore, once the CICFG of an app is constructed, we abstract it to obtain CADG and narrow down our focus only to its sensitive behaviors. 
	Intuitively, CADG of an app is obtained from its CICFG by considering only the nodes that access security-sensitive APIs\footnote{Two existing works, namely, PScout \cite{PScout} and SUSI \cite{Susi} list commonly known security-sensitive Android APIs. These two lists have been used to perform security and malware analysis by many existing approaches such as \cite{AppContext,Mudflow,ICCTA}. Following these studies, \tool{} uses these two lists to identify security-sensitive APIs.} along with their context information. All other nodes are removed and paths that exists between such nodes in the CICFG become edges in ADG, if they satisfy certain conditions as described below.
	
	\textbf{Definition 2 (CADG).} CADG can be represented as a 4-tuple $CADG = (N, E, \lambda, \xi)$, where $N$ is a finite set of nodes and $n \in N$ is an instruction of a method that corresponds to invoking a security-sensitive API and therefore $N \subseteq N_i$. $E \subseteq N \times N$  is a set of directed edges where an edge from $e (n_1, n_2) \in E$ exists iff there exists a path $p (n_1, n_2)$ between these two nodes in the CICFG and $ method(n_1) $ = $method (n_2) $, where $ method(n) $ denotes the method that encompasses instruction $n$\footnote{This follows from the observation that in most malware the malicious code portion is closely-knit i.e., spanning only to a few methods. We also attempted two other variants of CADG. We reduce the path in CICFG to edges in CADG (i) only if the calling and called nodes belong to the same package and (ii) only if the calling and called nodes belong to the same class. Both these variants contained much larger number of edges in CAGD and also failed to capture the attacks as effectively as the CADG defined above (experimentally verified).}. $\lambda$ is the set of labels representing the security-sensitive APIs and $\ell: N \rightarrow \lambda$ is a labeling function which assigns a label to each node. 
	%Hence, CADG is a directed node labeled graph. 
	$\xi$ is a set of triggers though which every node in the CADG could be reached and $\mathcal{C} \rightarrow \xi$ is a function which assigns the context to each node.
	
	A portion of \textit{Geinimi} and \textit{Yahoo Weather} examples' CADGs are shown in Fig. \ref{fig:why-cwlk} (c) and (d), respectively. In both these apps, all three CICFG nodes invoke security-sensitive APIs and they are retained in CADGs. All of these nodes belong to the same method (i.e. {\tt leak\_info\_to\_url}) and hence the path that existed among them in CICFG translate to edges in CADG.

	\subsection{Feature Extraction \& Representation using CWLK} 
	\label {ss:cwlk}
	
	Once the CADGs are constructed, our next task is to extract semantic features that characterize sensitive behaviors of apps from them. 
	%Every CADG node represents a sensitive API invocation.  Our method strives to capture sequences of such invocations and thus the neighborhood of a CADG along with the context in which a they are reached. 
	To this end, we use CWLK, a graph kernel we developed in our previous work \cite{cwlk} which is specifically designed to capture both structural and contextual information from PRGs and perform accurate malware detection. This directly addresses the requirements R1 and R2 stated in \S \ref{ss:mot-cwlk}. 
	
	We furnish the details on how CWLK supports \tool{} in extracting CADG subgraph features to perform malware detection, in this subsection. For detailed discussions on CWLK use cases with other PRGs (e.g., ICFG), comparison with WLK and other graph kernels, we refer the reader to \cite{cwlk}.
	%In summary, CWLK facilitates capturing the following information when extracting features from CADG: (1)	CADG neighborhood and (2) context of the central node of the neighborhood. 
	%To this end we use CWLK, a graph kernel we developed in our previous work \cite{cwlk}. For the sake of completeness, we furnish the details on how to apply CWLK to capture both structural and contextual information from PRGs and perform accurate malware detection here. For a detailed discussions on the use cases, comparison to other graph kernels, validity and time complexity, we refer the reader to \cite{cwlk}.
	%As it could be used with any PRG (e.g., CG, ICFG) for malware detection, we choose to explain how CWLK could be applied on PRGs in general rather than specifically discussing CADGs. 
	
	%We begin by explaining how the regular WLK can be applied to perform malware detection using PRGs and how it falls short. Subsequently, we introduce our CWLK and discuss how it addresses the shortcomings of WLK. Finally, we prove CWLK’s semi-definitiveness and analyze its time complexity.
	
	We begin by explaining the regular WLK, then introduce the proposed CWLK and finally discuss how WLK falls short when performing malware detection using CADGs and how CWLK rectifies the same.

	\textbf{Weisfeiler-Lehman Kernel.}
	WLK \cite{WLK}, works by decomposing graphs into subgraphs in such a way that a kernel function for a pair of graphs can be defined as a convolution of kernel functions defined over their subgraphs. %\textit{The main idea behind WLK is to condense the information contained in a PRG neighborhood subgraph into a single label value along with the context of the neighborhood}.
	
	WLK computes the similarities between a given pair of graphs $ G = (N, E, \lambda) $ and $ G' = (N',E',\lambda) $ based on the 1-dimensional WL test of graph isomorphism \cite{WLK}. The algorithm works by iteratively augmenting the node labels by the sorted set of labels of neighboring nodes. 
	This process is referred to as \textit{label-enrichment} and new labels are referred as \textit{neighborhood labels}. 
	Thus, in each iteration \textit{i} of the WL algorithm, for each node $n \in N$, we get a new neighborhood label, $\lambda_{i}(n)$ that encompass the $i^{th}$ degree neighborhood around $n$. 
	%$\lambda_{i}(n)$ could be optionally compressed using a hash function $f: \Sigma^* \rightarrow \Sigma$ such that $f(\lambda_{i}(n)) = f(\lambda_{i}(n'))$, iff $\lambda_{i}(n) = \lambda_{i}(n')$. 
	For graph $G$, this relabeling process yields a WL graph at height $i$, denoted as $\mathcal{G}_i$. Thus for any given graph $G$, we could obtain a sequence of WL graphs as defined below.
	
	\textbf{Definition 3 (WL sequence).} Define the WL graph at height $i$ of the graph $G = (N,E,\lambda)$ as the graph $\mathscr{G}_i = (N,E,\lambda_i)$. The sequence of graphs 
	\begin{equation}
	{\mathscr{G}_0, \mathscr{G}_1, ..., \mathscr{G}_h} = {(N,E,\lambda_0),(N,E,\lambda_1), ...,(N,E,\lambda_h)}
	\end{equation}
	is called the WL sequence up to height $h$ of $G$, where $\mathscr{G}_0 = G$ (i.e., $\lambda_0 = \lambda$) is the original graph and $\mathscr{G}_1 = r(\mathscr{G}_0)$ is the graph resulting from the first relabeling, and so on.

	The WL kernel over graphs $G$ and $G'$ leverages on their respective WL sequences and is defined as follows.
	
	\textbf{Definition 4 (WL kernel).} Given a valid kernel $k(.,.)$ and the WL sequence of graph of a pair of graphs $G$ and $G'$, the WL graph kernel with $h$ iterations is defined as 
	\begin{equation}
	\label{eq:wlk}
	k^{(h)}_{WL}(G, G') = k(\mathscr{G}_0, \mathscr{G}'_0) + ... + k(\mathscr{G}_h, \mathscr{G}'_h)
	\end{equation}
	where $h$ is the number of WL iterations and ${\mathscr{G}_0, \mathscr{G}_1, ..., \mathscr{G}_h}$ and ${\mathscr{G}'_0, \mathscr{G}'_1, ..., \mathscr{G}'_h}$ are the WL sequences of $G$ and $G'$, respectively. $ h $ is referred as \textit{height of the kernel}.
	
	Intuitively, WLK counts the common neighborhood labels in two graphs. 
	Hence we have $k^{(h)}_{WL}(G,G') = |(\lambda_i(n), \lambda_i(n'))|$, 
	%if $f(\lambda_i(n)) = f(\lambda_i(n'))$ for $i \in \{1, . . . , h\}, n \in N, n' \in N'$, 
	%iff $ f(\lambda_i(n)) = f(\lambda_i(n'))$ 
	for $i \in \{0, . . . , h\}, \forall n \in N, \forall n' \in N'$.
	%, where $f$ is injective and the sets $ \{f(\lambda_i(n))|n \in N\cup N'\} $ and $ \{f(\lambda_j(n))|n \in N\cup N'\} $ are disjoint for all $ i \neq j $.
	Clearly, this enables WLK to capture structural information from CADGs. However, there is no scope for capturing the reachability context information in WLK. This is exactly what we address through our CWLK.
	\nolinebreak
	%\textbf{Example \& WLK's shortcoming.}
	%We now apply WLK on the real-world examples discussed in \S \ref{ss:mot-cwlk} to see if it distinguishes malicious and benign neighborhoods clearly, facilitating accurate detection. 
	%%For the ease of illustration, the label compression step is avoided. 
	%Applying WLK on both \textit{Geinimi} and \textit{Yahoo Weather} CADGs (see Fig. \ref{fig:why-cwlk} (c) and (d)), for the node {\tt getLatitude}, for heights $h = 0, 1$, we get the neighborhood labels as shown in Fig. \ref{fig:illus} (a)-(d). 
	%%In both cases, the node {\tt getLatitude} has only one degree-1 neighbor, {\tt writeBytes} and this fact is reflected in the neighborhood label. 
	%Clearly, WLK captures the structural information around the node {\tt getLatitude}, incrementally in every iteration of $h$. 
	%In fact, neighborhood label for $h=1$ captures that another sensitive node, {\tt writeBytes} lies in the neighborhood of {\tt getLatitude}, which highlights a possible privacy leak. However, WLK does not capture whether the neighborhood involved in this leak is reached in \textit{user-aware} or \textit{unaware} context. Hence, whether or not the leak is malicious still remains inconclusive. This is precisely what we address through our CWLK.
	
	% In other words, the structural features extracted by WLK for both the benign and malicious sample are the same. A straight forward solution to this problem is to attach the context of the node {\tt getLatitude} its neighborhood labels.
	
	\begin{algorithm}[t]
		\small
		\caption{CWLK - Contextual re-labeling}
		\textbf{Input}:\\
		$G = G_0 = (N, E, \lambda_0, \xi)$ | \textit {CADG} with set of nodes ($N$), set of edges ($E$) and set of node labels ($\lambda_0$) and context for each node ($\xi$)\\
		$h$ | number of iterations\\
		\textbf{Output}:\\
		$\{\mathcal{G}_0, \mathcal{G}_1,...,\mathcal{G}_h\}$ - contextual WL sequence of height $h$ 
		\begin{algorithmic}[1]
			
			\Procedure{Contextual Re-label}{$G,h$}%\Comment{The g.c.d. of a and b}
			\For {$i = $ 0 to $h$}
			\For {\textbf{all} $n \in N$}
			\LState $\sigma_i(n) \leftarrow \emptyset$
			\If {$i = 0$}
			\For {$c \in \xi(n)$}
			\LState $\sigma_i(n) \leftarrow \sigma_i(n) \cup c \oplus \lambda_0(n) $ %\Comment{{\scriptsize Contextual Nei.hood label}}
			\EndFor
			\Else
			\LState $\mathcal{N}(n) \leftarrow \{m\ |\ (n,m) \in E\}$% \Comment {set of neighbors of n}
			\LState $M_i(n) \leftarrow \{\lambda_{i-1}(m)\ |\ m \in \mathcal{N}(n) \}$ 
			\LState $\lambda_i(n) \leftarrow  \lambda_{i-1}(n) \oplus sort(M_i(n))$  %\Comment{{\scriptsize Nei.hood label}}
			\For {$c \in \xi(n)$}
			\LState $\sigma_i(n) \leftarrow \sigma_i(n) \cup c \oplus \lambda_i(n) $ %\Comment{{\scriptsize Contextual Nei.hood label}}
			\EndFor
			
			\EndIf
			\LState $\gamma_i(n) \leftarrow string(\sigma_i(n))$ 
			%\LState $\gamma_i(n) \leftarrow f(\sigma_i(n))$ %\Comment{{\scriptsize Label compression}}
			\EndFor
			\LState $\mathcal{G}_i \leftarrow (N,E,\gamma_i)$ 
			\EndFor
			\LState \textbf{return} \{$\mathcal{G}_0,\mathcal{G}_1,...,\mathcal{G}_h$\}
			\EndProcedure
			
		\end{algorithmic}
		\label{algo:cr}
	\end{algorithm} 
	
	\textbf{Contextual Weisfeiler-Lehman graph Kernel.}
	\label{sss:cwlk}
	The goal of CWLK is to capture both structural and contextual information from PRGs. To this end, we modify the re-labeling step of WLK so as to accommodate the context of every neighborhood. We refer to this process as \textit{contextual-relabeling} and the sequence of graphs thus obtained as \textit{contextual WL sequence}.
	
	\textbf{Contextual re-labeling.}
	Specifically, CWLK performs one additional step in the re-labeling process which is to attach the contexts of every node to its neighborhood label in every iteration. This in effect, indicates the contexts under which a particular neighborhood is reachable. The label thus obtained is referred to as \textit{contextual neighborhood label}. 
	The contextual relabeling process is presented in detail in Algorithm \ref{algo:cr}.  
	\nolinebreak
	%The first input to algorithm is a CPRG $G$ with a set of nodes, edges, node labels and contexts $N$, $E$, $\lambda_0$ and $\xi$, respectively. The next input $h$ is the degree of neighborhoods to be considered for re-labeling. The output is the sequence of graphs $\{\mathscr{G}_0, \mathcal{G}_1,...,\mathcal{G}_h\}$ which contains same set of nodes and edges as $G$ and the original labels in $\lambda_0$ are replaced with a set of contextual neighborhood label for each node $\gamma_i$ in $\mathcal{G}_i$. The sequence of graphs $\mathcal{G}_i$ is referred to as \textit{contextual WL sequence}.
	\nolinebreak
	%\textbf{CWLK Re-labeling Algorithm.} 
	
	The inputs to the algorithm are CADG, $G$ and the degree of neighborhoods to be considered for re-labeling, $h$. The output is the sequence of contextual WL graphs, $\{\mathcal{G}_0, \mathcal{G}_1,...,\mathcal{G}_h\}\!=\!\{(N,E,\gamma_0), (N,E,\gamma_1),...,(N,E,\gamma_h)\}$, where $\gamma_1,...,\gamma_h$ are constructed using the contextual relabeling procedure.
	
	For the initial iteration $ i=0 $, no neighborhood information needs to be considered. Hence the contextual neighborhood label $ \gamma_0 (n) $ for all nodes $ n \in N $ is obtained by justing prefixing the contexts to the original node labels (lines 6-8,17). For $ i\!>\!0 $, the following procedure is used for contextual re-labeling.
	Firstly, for a node $n \in N$, all of its neighboring nodes are obtained and stored in $\mathcal{N}(n)$ (line 10). For each node $m \in \mathcal{N}(n)$ the neighborhood label up to degree $i-1$ is obtained and stored in multiset $M_i(n)$ (line 11). $\lambda_{i-1}(n)$, neighborhood label of $n$ till degree $i\!-\!1$ is concatenated to the sorted value of $M_i(n)$ to obtain the current neighborhood label,  $\lambda_i(n)$ (line 12). Finally the current neighborhood label is prefixed with the contexts of node $n$ to obtain $ \sigma_i(n) $ which is then represented as a string  $\gamma_i(n)$ which denotes the contextual neighborhood label (lines 13-15,17-18). 
	
	\textbf{Definition 5 (CWL kernel).} Given a valid kernel $k(.,.)$ and the CWL sequence of graph of a pair of graphs $G$ and $G'$, the contextual WL graph kernel with $h$ iterations is defined as 
	\begin{equation}
	\label{eq:cwlk}
	k^{(h)}_{CWL}(G, G') = k(\mathcal{G}_0, \mathcal{G}'_0) + ... + k(\mathcal{G}_h, \mathcal{G}'_h)
	\end{equation}
	where $h$ is the number of CWL iterations and ${\mathcal{G}_0, \mathcal{G}_1, ..., \mathcal{G}_h}$ and ${\mathcal{G}'_0, \mathcal{G}'_1, ..., \mathcal{G}'_h}$ are the CWL sequences of $G$ and $G'$, respectively. 
	
	Intuitively, CWLK counts the common contextual neighborhood labels in two graphs. 
	Hence we have $k^{(h)}_{CWL}(G,G') = |(\gamma_i(n), \gamma_i(n'))|$, 
	%iff $ f(\sigma_i(n))\!=\!f(\sigma_i(n'))$ 
	for $i \in \{0, . . . , h\}, \forall n \in N, \forall n' \in N'$.
	%, where $f$ is a label compression function as discussed above.\\
	
	\textbf{Illustrating WLK's shortcoming and CWLK's efficacy.}
	\begin{figure}[t]
		\centering
		\includegraphics[height=2.6cm,width=9cm]{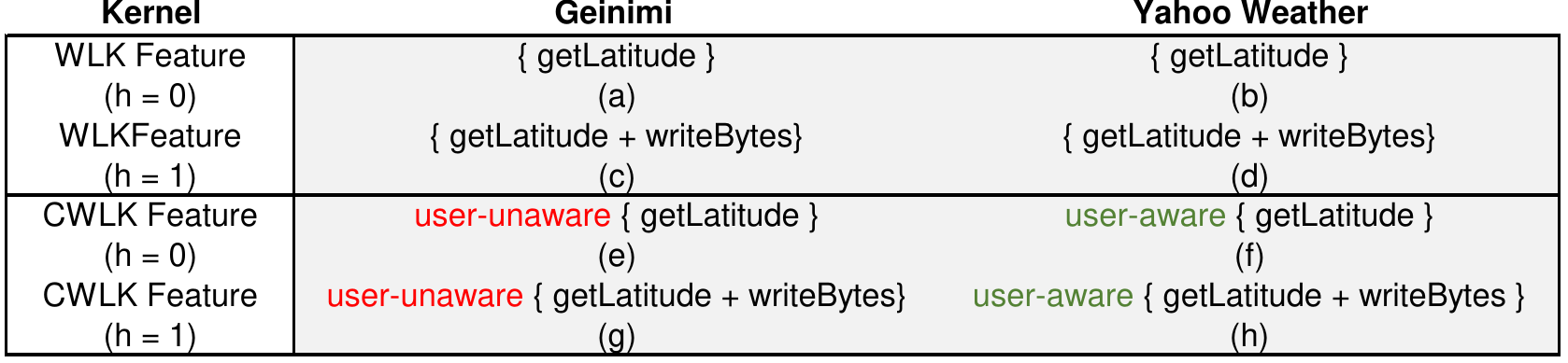}
		\caption{\small WLK and CWLK neighborhood labels for the node {\tt getLatitude} from \textit{Geinimi} and \textit{Yahoo Weather} apps. \label {fig:illus}}
	\end{figure}
	Having presented the formulations for both kernels, we now  illustrate WLK's shortcomings and explain how CWLK addresses the same, with an example.
	%First, we apply WLK on the  to see if it distinguishes malicious and benign neighborhoods clearly, facilitating accurate detection. 
	Applying WLK on the CADGs of \textit{Geinimi} and \textit{Yahoo Weather} examples  (see Fig. \ref{fig:why-cwlk} (c) and (d)), for the node {\tt getLatitude}, for heights $h = 0, 1$, we get the neighborhood labels as shown in Fig. \ref{fig:illus} (a)-(d). 
	In both cases, the node {\tt getLatitude} has only one degree-1 neighbor, {\tt writeBytes} and this fact is reflected in the neighborhood label. 
	Clearly, WLK captures the structural information around the node {\tt getLatitude}, incrementally in every iteration of $h$. 
	In fact, neighborhood label for $h=1$ captures that a sensitive node, {\tt writeBytes} lies in the neighborhood of {\tt getLatitude}, which highlights a possible privacy leak. However, WLK does not capture whether the neighborhood involved in this leak is reached in \textit{user-aware} or \textit{unaware} context. Hence, whether or not the leak is malicious still remains inconclusive. This is precisely what CWLK addresses. 
	On applying CWLK for the same node in both apps, we obtain the contextual neighborhood labels as shown in Fig. \ref{fig:illus} (e)-(h).
	Clearly, the contextual neighborhood labels of \textit{Geinimi} reveal that the sensitive operations are performed in the \textit{user-unaware} context, whereas, the same for \textit{Yahoo Weather} happen in the \textit{user-aware} context.
	Hence, it is evident that the CWLK’s contextual relabeling provides a means to clearly distinguish malicious CADG neighborhoods from the benign ones. 
	%This is achieved by complementing the structural information with contextual information. 
	Therefore, unlike WLK, CWLK based classification does not detect \textit{Yahoo Weather} as a false positive. This example clearly establishes the suitability of CWLK for the malware detection task.

	\textbf{CWLK validity and complexity.} CWLK's positive definiteness is asserted in Appendix \ref{app:mercer}. The runtime complexity of CWLK with $h$ iterations on a graph with $n$ nodes and $e$ edges is $O(he)$ which is same as that of WLK. Meaning, capturing contextual information in CWLK does not reduce the efficiency. For derivation and discussion on time complexity, see Appendix \ref{app:tc}.
	
	\textbf{Efficient computation of CWLK on K graphs.} When computing CWLK on $K$ graphs to obtain $K \times K$ kernel matrix, a naïve approach would involve $K^2$ comparisons, resulting a time complexity of $O(K^2he)$. However, as mentioned in \cite{WLK}, a BoF model based optimization could be performed to compute the kernel matrix in $O(Khe+K^2hn)$ time. This optimized computation involves the following steps: (1) A vocabulary $\Sigma^*$ of all the contextual neighborhood labels of nodes across the $K$ graphs is obtained in $O(Khe)$ time. This facilitates representing each of the $K$ graphs as feature vectors of $|\Sigma^*|$ dimensions. (2) Subsequently, $K \times K$ kernel matrix can be computed by multiplying these vectors in $O(K^2hn)$ time. 
	In summary, CWLK has the same efficiency as WLK yet supports richer feature vector representations which retain CADGs'  expressiveness better.
	
	%\textbf{Relation to other spatial contextual kernels.}
	%Two recently proposed graph kernels \cite{ContextualKernel} and \cite{NSPDK}, consider incorporating the spatial context information to neighborhood subgraph features. They define \textit{context of a subgraph feature as another subgraph appearing in its vicinity}.
	%As mentioned earlier, in our malware detection problem we refer to \textit{attributes of a node which determine its reachability as its context}. 
	%This \textit{reachability context} is different from \textit{spatial context} discussed in \cite{ContextualKernel} and \cite{NSPDK}. Hence CWLK is consummately different from these two kernels.

	\subsection{Online Learning}
	\label{ss:ol}
	Once the feature vectors of all the apps in the training-set are built using CWLK, we train an online CW \cite{cw} classifier with them to detect malware. CW classifier's training and update procedures are as explained below with relevant notations.
	
	Denote the features of an app (both benign and malware) as a vector $x = [x^{(1)}, x^{(2)},...,x^{(d)}]^{T}$, %where $x^{(j)}$ denotes individual features 
	and its label as $y \in \{-1, +1\}$, where $-1$ indicates benign and $+1$ indicates malicious apps. The CW classifier receives a number of samples, $x_i$, and their labels, $y_i$, and trains using this labeled data. Given a new unseen sample, $x$, the goal of CW classifier is to predict the label, $y$, of this new sample based on its trained model. 
	
	CW classifier fits a linear decision boundary (i.e., hyperplane) between the positive and negative class samples. That is, the model is a weight vector, $w = [w^{(1)}, w^{(2)},...,w^{(d)}]^{T}$ which indicates the weight (i.e., relative importance) of each of the features used to predict the output label $y$. The predicted label, $\hat y$, is the sign of the inner product between $x$ and $w$:
	\begin{equation}
	\hat y = sign (x \cdot w)
	\end{equation}
	
	CW incrementally builds the model in eq. (4) in rounds. In round $t$, it receives a sample, $x_t$ and predicts its label $\hat {y_t}$ using the current model; it then receives $y_t$, the true label of $x_t$ and updates its model based on the sample-label pair: $(x_t, y_t)$. 
	%That is, $w$ is updated if the predicted label, $\hat {y_t}$ and the true label, $y_t$ of the sample $x_t$ are not the same.
	%
	%The goal of the CW algorithm is to update the model $w$ as minimal as possible to correct for any mistakes it commits. PA solves the following optimization with each given sample: 
	%\vspace{-2mm}
	%\begin{equation}
	%w_{t+1} \leftarrow \operatornamewithlimits{argmin}\limits_{w} \frac{1}{2} || w_t - w ||^2
	%\end{equation}
	%\vspace{-6mm}
	%\begin{center}
	%	subject to $y_i (w \cdot x_t) \geq 1$
	%\end{center}
	%Updates occur only when  $y_t(w_t \cdot x_t) < 1$., The closed-form update for all samples is as follows:
	%\begin{equation}
	%w_{t+1} \leftarrow w_t + \alpha_ty_tx_t
	%\end{equation}
	%where $\alpha_t = max\{\frac{1 - y_t(w_t \cdot x_t)} {||x_t||^2} , 0\}$ (we refer the reader to the original work at \cite{PA} for this derivation and further details on PA algorithm). 
	%%The PA algorithm has been successful in practice because the updates explicitly incorporate the notion of classification confidence.
	A Gaussian distribution over the weights with mean $\mu$ and covariance matrix $\Sigma$ is maintained by the CW algorithm. The value $\mu^{(f)}$ represents mean weight of feature $f$ (i.e., mean of $w^{(f)}$), and the value $\Sigma_{f,f}$ captures the confidence in $f$'s weight. While classifying a new sample $x$, the weight $w$ is drawn from $N (\mu, \Sigma)$. In practice, one could choose $w = \mu$, the average weight vector and use eq (4) to arrive at the predicted label. CW updates the model, (i.e., $\mu$ and $\Sigma$), continuously on every labeled sample instead of only when committing misclassifications. The rationale is that \textit{making correct prediction also suggests that the model should increase its confidence of the current weights}. CW's update rule is presented below: 
	\begin{equation}
	(\mu_{t+1}, \Sigma_{t+1}) = arg \min_{\mu, \Sigma} D_{KL}(\mathcal{N}(\mu, \Sigma)||\mathcal{N}(\mu_t, \Sigma_t)), 
	\end{equation}
	\begin{equation}
	s.t. Pr_{w \sim N(\mu,\Sigma)}[y_t(w \cdot x_t)] \ge \eta .
	\end{equation}
	Eq. (5) determines that the new distribution characterized by new $\Sigma$ and $\mu$ should be close to the old distribution as much as possible. The KL divergence ($D_{KL}$) provides the measure of distance between the two distributions. Eq. (6) determines that the update should ensure that the probability of making correct prediction if the same sample, $x_t$ is seen in the next round must be bigger than $\eta$, where $\eta$ is a configurable parameter (usually set bigger than 50\%). The computational complexity of the update is linear in the number of non-zero features in $x_t$. %The memory required is constant in the input sample, i.e., the memory for the current $x$. 
	%We refer the reader to \cite{cw} for more details on CW. 
	
	The strength of CW lies in its notion of modeling confidence on features' weights. Evidently, if the weight of a feature does not fluctuate very much over time, one could confidently believe that this weight is what it should be. CW achieves the two following desirable characteristics through this confidence notion, which are not exhibited by other online learners (e.g., Online Perceptron (OP), Passive Aggressive (PA)):
	
	1. The weights of more confident features are updated less aggressively. For instance, using {\tt sendSms} API in the \textit{user-unaware} context is a good indicator of an app's maliciousness; consequently, its weight does not get updated abruptly over time, thereby instigating high confidence on this feature. Therefore, CW ensures that the weight will not change much even when it receives an instance of benign app using \textit{user-unaware}\{{\tt sendSms}\}, which possibly could be a case of incorrect labeling. This makes CW naturally robust to labeling noise.
	
	2. The model is updated just enough to adapt to the changing trends in the data while refraining from changing too much.
	The rationale is that the previous model carries valuable information about the data and should not be modified too abruptly.

	\textbf{Alternatives.} As practitioners involved in building an online malware detection framework, we do not have any vested interest in a particular algorithm for online learning. Ultimately, we wish to determine the algorithm that scales well to problems of our volume and complexity and yields the best performance. 
	To that end, we experimented with other well-known online learning algorithms such as OP \cite{op} and Stochastic Gradient Descent learning based Logistic Regression (LR\textsubscript{SGD}) \cite{lrsgd} and PA \cite{PA}, along with CW. Since CW offered the best results in terms of accuracy and efficiency, it is used in \tool. 
	
	We believe this is because OP and LR\textsubscript{SGD} do not imbibe the notion of classification confidence and treat all misclassifications equally. PA, on the other hand, updates more aggressively when the margin of error is large and less aggressively, otherwise. Meaning, even though PA imbibes confidence of the classification decision, it does not account for the confidence over individual features. This leads PA to vary the weights of features rapidly and thereby making it susceptible to fitting to noise. On the contrary, CW imbibes both the aforementioned confidences and thus is robust to noise in the stream. 
	%We also experimented with newer online learners like Confidence Weighted (CW) \cite{cw}, Soft Confidence Weighted (SCW) \cite{scw} and Adaptive Regularization of Weight Vectors (AROW) \cite{arow} algorithms. 
	
	\textbf{Online Kernel Classifiers.} We experimented with online kernel (i.e., nonlinear) classifiers such as Projectron \cite{projectron} and Forgetron \cite{forgetron} (with budget). 
	Since these algorithms offered marginal improvement in accuracy at a significantly worse efficiency in our large-scale dataset, we chose CW to cater \tool's online learning requirements.
	
	Overall, as the CW algorithm makes a fine-grained distinction between each feature's weight confidence, it can be especially well-suited to detect malware apps since our data stream continually introduces a dynamic mix of new and recurring CADG subgraph features. 
	Once the CW classifier in \tool{} is trained with all the samples it is ready to perform malware detection at scale. %It is important to note that since \tool{} is trained in an online fashion, it performs malware detection and simultaneously adapts to the changing trends in malware features by retraining on every sample it misclassifies.
	
	%\subsection{Explanation}
	%ML based malware detection solutions are usually black-box methods as they do not explain why a particular sample is detected as malicious or benign. In the case of DroidOL, we address this shortcoming as follows: Besides detection \tool{} reports significant CADG subgraphs of an app that contribute to a detection. Since \tool{} uses a linear classifier (i.e., PA) with interpretable features (rooted subgraphs), we are able to determine the contribution of each CADG subgraph feature to the final class prediction. As in eq. X in section XX, during the prediction of ˆy, a given sample’s class, the largest k weights w (s) which are significantly important for placing the sample on malicious side of the decision hyperplane are identified. Since each weight w (s) is assigned to a certain feature x (s) , it is then possible to explain why an app has been classified as malicious or not. After extracting the top k CADG graphlet features, \tool{} reports them. This explains the significant semantic characteristics that makes a particular sample malicious or benign. From our evaluations, we observe that more often than not, the malware features include CADG subgraphs representing use of dynamic/reflection code, sending of SMS and internet communication in the user-unaware context. The detailed procedure used for explaining DroidOL’s predictions are presented in section XX.
	
	\subsection{Explaining \tool's predictions}
	\label{ss:exp}
	
	Once \tool{} classifies a sample as malicious or benign, the next step is to offer explanations of these predictions by reporting significant CADG subgraphs of an app that contribute to prediction. These explanations help to understand whether the rationale behind the model's predictions are reasonable, thus ensuring users' trust on the model. Besides, they could help human analysts in several tasks such as studying malware attacks and creating malware signatures. 
	%One will not deploy a malware detection model unless it is trustworthy and explanations help in assessing the trust. 
	
	Explaining the predictions of linear classifiers is a well-studied problem \cite{Drebin,Adagio,lime}. Following them, we are able to determine the contribution of each CADG subgraphs feature to the final class prediction as described below. 
	
	Based on eq. (4) in section \S \ref{ss:ol}, for a given sample $x$, during the prediction of $\hat y$, the largest $\nu$ weights $w^{(s)}$ which are significantly important for placing the sample on the malicious (or benign) side of the decision boundary are identified (i.e., a point-wise multiplication $wx = w \times x$ is performed and  $\nu$ largest values are obtained from the resulting vector $wx$). Since each weight $w^{(s)}$ is assigned to a certain feature $x^{(s)}$, it is then possible to explain why an app has been classified as malicious (or benign). After extracting the top $\nu$ CADG subgraph features, \tool{} reports them explaining the significant semantic characteristics that make a particular sample malicious (or benign). As we could observe from our evaluations, the frequently observed malware features include CADG subgraphs representing the use of reflection, dynamic and native code, accessing private information, sending of SMS and communication over internet predominantly in the \textit{user-unaware} or \textit{unresolved} context.

	%\subsection{Explainable Detection}
	%\label{ss:ed}
	%It has to be noted that not all classifiers are interpretable in their original feature space. For instance, the results of SVMs using RBF kernels, MLP neural networks stem from their ability to project to a high-dimensional feature space and learn complex non-linear decision boundaries. Hence, their prediction results could not be interpreted in the original input space. On the other hand linear and quasi-linear classifiers (e.g., linear SVM, PA, RF etc.) are known for their ability to provide interpretable results. Since our goal is to provide interpretable results, we make use of linear online leaner in the \tool{} framework. This makes \tool’s results interpretable.
	
	\section {Experimental Design \& Implementation}
	\label{sec:ed}
	We conducted several large-scale experiments to evaluate \tool's accuracy, efficiency, adaptiveness and explainability which are its primary design goals. We also compare its performance against two accurate and efficient state-of-the-art malware detection approaches. 
	%The experimental design (\S \ref{ss:exp_des}) and results and discussions (\S \ref{ss:r_d}) are presented in the two following sub-sections.
	In this section, experimental design aspects such as research questions (RQs) addressed, datasets used, evaluation setup and metrics are presented along with implementation details.
	
	\subsection {Research Questions}
	We intend to address the following RQs through our evaluations:
	
	\textbf{(RQ1 Accuracy)} \textit{How accurate is \tool{} in detecting unseen malware from both benchmark and real-world datasets?} %More specifically, we investigate the impact of two factors: (1) including contextual information through the proposed CWLK and (2) performing online learning over batch learning?
	
	The impact of the two most important factors responsible for \tool's accuracy are investigated separately through two following sub-RQs.
	
	(RQ1.1) What impact does capturing contextual information through CWLK have on \tool's accuracy?
	
	(RQ1.2)  Does \tool's online learning provide any benefit over batch learning in terms of accuracy?
	
	\textbf{(RQ2 Efficiency)} \textit{How efficient is \tool{} over batch learning based solutions in terms of overall training and prediction time?}
	%Again, we investigate the two following factors: (1) does including the context information through CWLK adversely affect the efficiency and (2) does the use of online learning improve the efficiency significantly?
	
	\textbf{(RQ3 Explainability)} \textit{How explainable are \tool’s predictions?} %In other words, does it provide reasons for classifying a sample as malicious or benign, rather than acting as a ‘black box’?
	
	\textbf{(RQ4 Adaptiveness)} \textit{How adaptive is \tool{} when performing malware detection in the real-world setting?} In other words, does it adapt to malware population drift seamlessly?
	
	\subsection{Datasets}
	\label{ss:ds}
	Many existing solutions such as \cite{DroidSift,AppContext,Mudflow,Drebin} are evaluated only using benchmark datasets. We observe that malware in benchmark datasets are much easier to detect compared to the ones found in-the-wild (as reported in \cite{AcsacEmpEval, CSBD}). Hence we evaluate \tool{} on both benchmark and a large collection of recent real-world malware collected in-the-wild, so as to exhibit its potential as a practical real-world malware detection solution. The details of these two datasets are presented below.
	
	\textbf{Benchmark (BM) dataset.} \textsc{Drebin} \cite{Drebin} provides a popular benchmark dataset comprising 5560 malware samples belonging to 179 families. This dataset is used in our evaluation. The date of creation of these apps fall in the range: from Mar'09 to Oct'12. The \textsc{Drebin} collection forms the malware portion of the dataset and for the benign portion, we used 5000 randomly chosen benign apps from Google Play \cite{Play} that are created in the same period. 
	%Thus our BM dataset consists of 5560 malware and 5000 benign apps that are created from 2009 to 2012. 
	
	\textbf{In-the-wild (ITW) dataset.} We collected a dataset of 44,347 benign and 42,910 malware apps from seven different markets in 2014. We infer that the date of creation of these apps fall in a span of 224 days starting from 1 Jan'14 to 13 Aug'14. Table \ref{tab:ds} provides information on the distribution of the applications in our dataset along with names of the markets where they have been crawled from. 
	Following the software security research practices proposed in \cite{Drebin} and \cite{CSBD}, we leveraged on the VirusTotal web portal\footnote{\url{https://www.virustotal.com}} to infer the ground truth labels for all the apps in the ITW dataset.
	
	%\textbf{Ground Truth Labels.} The apps in the BM dataset are labeled to be malicious or benign, precisely, following well-known software security research practices \cite{Drebin}. However we need such ground-truth labels for the apps in the ITW dataset to train ML models. Hence, following the strategy proposed in \cite{Drebin, CSBD}, we use the VirusTotal web portal\footnote{\url{https://www.virustotal.com}} which hosts malware detection services from more than 40 AV scanners to determine the ground-truth labels of the apps in ITW dataset. An app is labeled as malware if at least two of these scanners flags it as such. 
	
	\begin{table}[t]
		\centering
		\scriptsize
		\caption{{In-the-wild Dataset with apps dated from Jan. 1 2014 to Aug. 13 2014}}
		\label{tab:ds}
		\begin{tabular}{|l|c|c|}
			\hline
			{\bf Market} & {\bf \# of Benign Apps} & {\bf \# of Malware} \\ \hline
			Google Play\cite{Play}  & 39156                   & 26178                \\ \hline
			Anzhi \cite{AnZhi}        & 2957                    & 12260                \\ \hline
			AppChina \cite{AppChina}     & 1845                     & 4154                \\ \hline
			SlideMe \cite{SlideMe}      & 289                     & 132                  \\ \hline
			HiApk \cite{HiApk}        & 65                      & 157                  \\ \hline
			FDroid \cite{FDroid}       & 29                      & 2                   \\ \hline
			Angeeks \cite{Angeeks}      & 6                       & 27                  \\ \hline
		\end{tabular}
		
	\end{table}

	\subsection{Experimental Setup.} 
	All the experiments were conducted on a server with 36 cores of Intel(R) Xeon CPU E5-2699 v3 @ 2.30GHz and 32 GB RAM running Ubuntu 14.04.
	
	\subsection {Implementation and Comparative Analysis}
	\label{ss:impl}
	\tool{} is implemented in approximately 17200 lines of Python and Java code. IccTA \cite{ICCTA} an Android static analysis workbench based on Soot \cite{Soot} (which encompasses IC3 \cite{IC3} for ICC analysis) is used for building CADGs from apps. Scikit-learn \cite{SKLearn} toolbox is used for \tool's ML functionalities. 
	
	\textbf{Comparison with \soa{} solutions.} \tool{} is compared against two state-of-the-art Android malware detection solutions, namely, \textsc{Drebin} \cite{Drebin} and Allix \textit{et al.} \cite{CSBD}. To this end, we re-implemented these two approaches. Since the malware detection accuracy of these solutions predominantly depend on the features they use, we briefly introduce them here. 
	
	\textbf{\textsc{Drebin}} \cite{Drebin} is well-known for its scalable and explainable detection. It extracts light-weight semantic features such as APIs and permissions used, URLs accessed, names of components from apps and subsequently, trains a linear SVM classifier to distinguish malware from benign apps. 
	
	\textbf{Allix et al.} \cite{CSBD} recently proposed another scalable approach using structural features, namely Control Flow Graph (CFG) signatures. Therefore, we refer to this technique as CFG-Signature Based Detection (CSBD) in the reminder of the paper. 
	CSBD constructs CFGs of individual methods and encodes them as text-signatures. Subsequently, a RF classifier is trained with these signatures to detect malware.
	
	We re-implemented \textsc{Drebin} and CSBD in 1400 and 900 lines (approx.) of Python code, respectively. 
	Authenticity and correctness of our re-implementations is verified as we observe their accuracy and scalability values very similar to the ones reported in the original work on similar experiments (see \S \ref{sec:r_d}). 
	Besides this, re-implementations have been done in consultation with the authors of original work.
	
	%{\color {violet} It is noted that tools such as Soot and IccTA which are used in the feature extraction phase of the compared solutions, fail to process some \textit{apk} files due to runtime errors (eg. errors in packaging \textit{apk} files). Hence, this along with other factors such as timeouts and insufficient memory allocations results in failures, inducing a difference in number of apps processed by each solution in the following experiments.}

	\subsection{Evaluation metrics.}
	Standard evaluation metrics such as Precision, Recall, F-measure and cumulative error rates are used to determine the effectiveness of malware detection. All these values are in the range [0, 1]. High values of precision, recall and F-measure and low values of cumulative error rate indicate accurate detection. Efficiency is determined in terms of time required to train and test the classifiers. Lower values of training and testing durations indicate efficient detection. 
	%Details of the aforementioned metrics could be found at \cite{MLBook}.
	
	\section {Evaluation}
	\label{sec:r_d}
	The evaluations, results and discussions pertaining to each of the RQs are presented in this section.
	
	\subsection{RQ1: Accuracy}
	\label{ss:rq1}
	
	The results for CWLK's and online learning's impact on \tool's accuracy are presented in the two following subsections.
	
	\subsubsection{\textbf{(RQ1.1) Impact of CWLK}}
	\begin{table*}[t]
		\centering
		\scriptsize
		\caption{Comparison of CWLK against state-of-the-art kernel and malware detection models - avg. ($\pm$ std) over 5 runs}
		\label{tab:acc1}
		\begin{tabular}{|c|ccc|ccc|c|c|}
			\hline
			& \textbf{\begin{tabular}[c]{@{}c@{}}WLK+\\ SVM(h=0)\end{tabular}} & \textbf{\begin{tabular}[c]{@{}c@{}}WLK+\\ SVM(h=1)\end{tabular}} & \textbf{\begin{tabular}[c]{@{}c@{}}WLK+\\ SVM(h=2)\end{tabular}} & \textbf{\begin{tabular}[c]{@{}c@{}}CWLK+\\ SVM(h=0)\end{tabular}} & \textbf{\begin{tabular}[c]{@{}c@{}}CWLK+\\ SVM(h=1)\end{tabular}} & \textbf{\begin{tabular}[c]{@{}c@{}}CWLK+\\ SVM(h=2)\end{tabular}} & \textsc{\textbf{Drebin}\cite{Drebin}} & {\textbf{CSBD}\cite{CSBD}} \\ \hline
			\textbf{P} & 97.11($\pm$0.35)  & 98.50($\pm$0.22) & 99.21($\pm$0.08) & 98.86($\pm$0.26) & 99.07($\pm$0.09) & \textbf{99.56($\pm$0.04)} & 99.01($\pm$0.20) & 98.61($\pm$0.46) \\ %\hline
			\textbf{R} & 95.27($\pm$1.21)  & 97.11($\pm$0.51) & 98.08($\pm$0.32) & 92.79($\pm$0.98) & 97.88($\pm$0.47) & 98.90($\pm$0.28) & \textbf{99.15($\pm$0.08)} & 99.13($\pm$0.12) \\ %\hline
			\textbf{F} &  96.18($\pm$0.63)  & 97.80($\pm$0.16) & 98.64($\pm$0.29) & 95.73($\pm$1.05) & 98.47($\pm$0.17) & \textbf{99.23($\pm$0.11)} & 99.08($\pm$0.09) & 98.87($\pm$0.25) \\ \hline
		\end{tabular}
	\end{table*}
	In order to determine whether incorporating both structural and contextual information improves \tool's accuracy, we study the two following scenarios: (1) use only the structural information from CADGs, (2) use both structural and contextual information from CADGs to perform malware detection. Understandably, the former scenario is realized through using WLK and the latter is realized through CWLK. Hence in this experiment, we compare the accuracies of these two kernels on our task.
	
	We experimented with different kernel heights, $h = $ 0 to 5 for CWLK and WLK (see eq. (2) and eq. (3)). The average number of (contextual) neighborhood features are found to increase exponentially with increasing values of $h$ as long as $h\!\le \!2$. However, this number does not increase significantly after $h$ = 2. This is because we have removed nodes that do not access sensitive APIs, which affects the connectivity and restricts CADG neighborhood sizes. In other words, we seldom have neighborhoods around nodes that span beyond degree 2 in CADGs. Similar trend is observed in WLK. Hence we restrict the height $ h $ to be 0, 1 and 2 for both the kernels. Thus, in our experiments, for each kernel, we have three malware detection models (one for each value of $ h $). \
	
	%\textbf{Classifier.} 

	\textbf{Dataset \& Experiments.} 
	The BM dataset is used in this experiment. 70\% of the samples (chosen at random) are used to train and the remaining 30\% samples are used to test the models' performances.
	The classifiers' hyper-parameters are determined on the training set using 5-fold cross-validation, whereas the test set is only used for determining the final detection performance. We repeat this procedure 5 times and average the results.
	Since we wish to exclude the impact of online learning on the models' accuracy, we use a canonical batch kernel classifier, namely SVM with both the kernels. The models are referred as CWLK+SVM and WLK+SVM, denoting the kernel and the classifier, respectively. In order to study how CWLK features compare to state-of-the-art solutions, \textsc{Drebin} \cite{Drebin} and CSBD \cite{CSBD} are included in this comparative analysis. The precision, recall and F-measures of these models are compared.

	%\textbf{Compared Methods.} 
	
	\textbf{Results \& Discussion.}
	These results for the 8 malware detection models under comparison are presented in Table \ref{tab:acc1}. The following inferences are drawn from the table:
	\begin{itemize}[leftmargin=*]
		\setlength\itemsep{0em}
		
		\item At the outset, for both CWLK and WLK, considering larger neighborhoods helps capturing the structural information better which in turn reflects in better performances. This is evident as Precision, Recall and F-measure values get better with increasing values of $ h $ for both the kernels. {\color{black}{However, this observation may not hold for large values of $ h $, as nodes that are far apart will be considered for neighborhood re-labeling, leading to a noisy re-labeling process.}}
		
		\item It is clear that CWLK outperforms WLK for significant values of $h$ (i.e., 1 and 2), in terms of F-measure. \textit{Since the only difference between WLK and CWLK is the latter’s capability to capture the context information, evidently, this is the reason for CWLK’s superior performance}.
		
		\item {\color{black}{Also, CWLK achieves better Precision than WLK for all values of $h$, consistently. This indicates that CWLK suffers lesser false positives than WLK.}} This reduction is a direct result of capturing context information which helps to precisely distinguish malicious CADG neighborhoods from the benign ones. 
		
		\item Since CWLK uses both contextual and structural information, it is important to analyze the contribution of each of these types of information to its performance. 
		Capturing only structural information is equivalent to using WLK. Hence from columns 1 to 3 of Table \ref{tab:acc1}, it is evident that structural information alone could provide a minimum of 96.18\% and an average of 97.54\% F-measure. 
		Similarly, the contribution of contextual information alone is ascertained using CWLK and setting $h=0$ to be 95.73\% F-measure. 
		Finally, the effect of using both types of information is ascertained by using CWLK and setting $ h>0 $ to be a minimum of 98.47\% and an average of 98.85\% F-measure. This clearly conveys that structural information is primary for performing effective malware detection and contextual information complements it, thereby helping to improve the accuracy. % CWLK attains superior performance by capturing both these types of information.
		
		\item When comparing CWLK to the state-of-the-art solutions, CWLK (with $ h=2 $) outperforms all the compared solutions in terms of F-measure and Precision. In particular, it outperforms the second best performing technique (i.e., \textsc{Drebin}) by 0.15\% F-measure. In terms of Recall, it outperforms CSBD and is comparable to \textsc{Drebin}. 
		%In terms of Precision, ours outperforms other methods.
		% Even though the absolute values of difference between CWLK and \textsc{Drebin} are marginal, considering the range (99-100\%), the improvement achieved by CWLK is significant.

		%\item From the feature engineering view-point, \textsc{Drebin} does not use both structural and contextual features. CSBD use structural information but not contextual information. This reveals that capturing both these types of information is the reason for our approach's superior performance, reinforcing our finding mentioned in point 2.
		
	\end{itemize}

	From the results and discussions above demonstrate CWLK's suitability for the malware detection problem. Hence, we consider using CWLK in all the remaining experiments unless mentioned otherwise.
	
	\subsubsection{\textbf{(RQ1.2) Impact of Online learning}}
	\label{ss:why-ol}
	\begin{figure}[t]
		\centering
		\includegraphics[height=5cm,width=9cm]{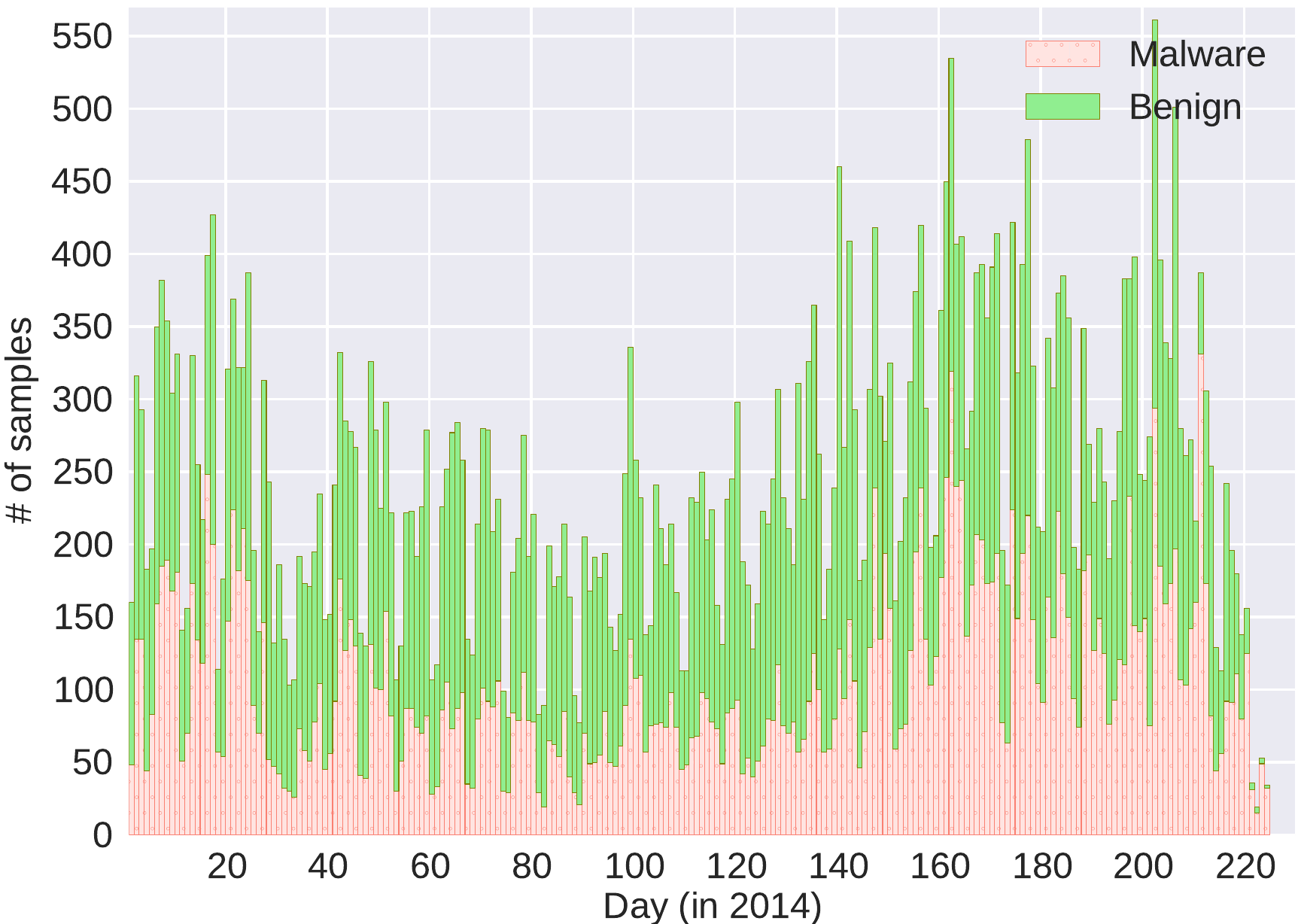}
		\caption{\small Timeline based distribution of apps in our large-scale ITW dataset (with malware and benign apps proportions). \label {fig:ds_hist}}
	\end{figure}
	
	\begin{figure*}[t]
		\centering
		\includegraphics[height=6.5cm,width=18cm]{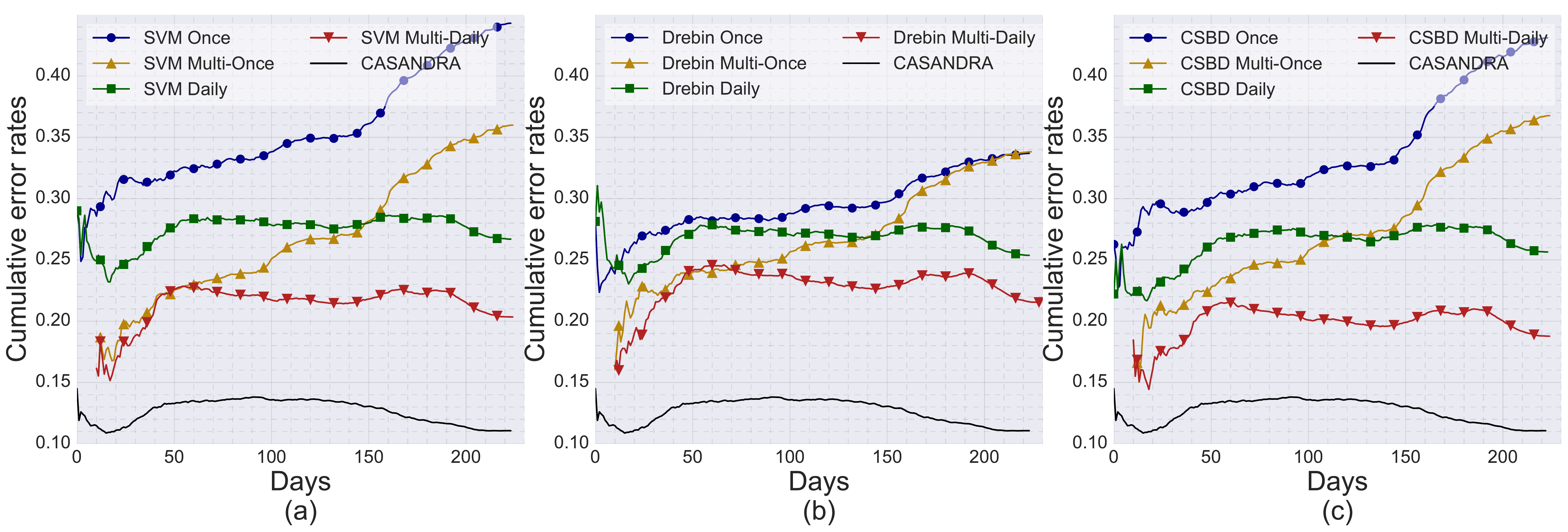}
		\caption{\small \tool: : Cumulative error rates for \tool{} vs. batch learning algorithms (a) \tool{} Vs. four variants of SVM, (b) \tool{}
			Vs. four variants of \textsc{Drebin}, (c) \tool{} Vs. four variants of CSBD. \label {fig:why-ol}}
	\end{figure*}
	
	With the inference on CWLK's impact arrived at, we now intend to evaluate the benefit of using online over batch learning for malware detection.
	
	\textbf{Dataset \& Experiment.} In this experiment, we use the ITW dataset as it is collected in the wild and larger, thus offering more scope for illustrating malware population drift.  We divide the ITW dataset apps (both benign and malware) into batches according to their date of creation.
	The resulting time-line based distribution of the two datasets are presented in Fig. \ref{fig:ds_hist}. We have 224 batches, one for each day of the ITW collection. 
	
	%\textbf{Compared Batch Learning Methods.} 
	Specifically, in this experiment, we intend to illustrate the benefits that \tool{} attains due to its use of online learning. To study this impact, we replaced the online learner in \tool{} with a canonical batch learner, namely, SVM and compare the same with \tool{} under different experimental settings (which involve periodic retraining). The above-mentioned SVM variants use the same features as \tool{} and hence it is sufficient to compare online and batch learning paradigms. %However, this does not reflect the significance of \tool{} as a practical malware detector in the context of current state-of-the-art malware detection techniques. In order to study this, we compare \tool{} with the state-of-the-art malware detectors, \textsc{\textsc{Drebin}} \cite{Drebin} and CSBD \cite{CSBD}.
	Furthermore, in order to compare \tool{} to \soa{} solutions, like in the previous RQ, we include \textsc{Drebin} and CSBD (which use different set of features) in this comparative analysis.
	
	\textbf{Batch Learning Configurations.}
	For batch-learning solutions, the classifiers are trained/retrained in a sliding window fashion over the 224 batches as explained below. For SVM, we experiment with the following variants: SVM-Once, SVM-Daily, SVM-MultiOnce, and SVM-MultiDaily. For SVM-Once, SVM classifier is trained only once on the apps from Day 1 (1 Jan’14). This model is tested on all other days without retaining. For SVM-Daily, the classifier is retrained after every day; however, only one previous day’s samples are used for every retraining — e.g., 11 Jan’14 results reflect training on the apps created on 10 Jan’14, and testing on 11 Jan’14 apps. SVM-MultiOnce is similar to SVM-Once and SVM-MultiDaily is similar to SVM-Daily, however, with the size of the batch for training and retraining covers 10 days instead of 1. In summary, for Once and MultiOnce variants, the model is never retrained and for Daily and Multi-Daily, the model is retrained in a sliding window fashion over the batches of data. The size of MultiDaily training sets is determined to be 10-day batches based on the data that our evaluation machine can handle.
	
	\textbf{\tool's online training regimen.}
	Since \tool’s feature extraction using CWLK is based on BoF model, its number of features grows as the samples stream in. Fig. \ref{fig:feat_growth} (in Appendix \ref{app:fg}) shows the cumulative number of features for each day of the evaluation in ITW dataset, representing the feature space growth. Each day’s total includes new features introduced that day and all the old features from previous days. We obtained a total of 15,171 features from the samples encountered on Day 1 (1 Jan'14). The dimensionality grows quickly as we extract new subgraph features from samples encountered every day and while reaching the final day (13 Aug'14), we have accumulated 2,114,050 features.
	This phenomenon of feature space growth is common across many techniques including \textsc{Drebin} and CSBD. This is because apps evolve over time for various reasons such as capability enhancements, bug fixes and adapting to changes in Android framework APIs \cite{Drebin,AcsacEmpEval,prescience}. This evolution results in newly observed characteristics which translate into new features from an ML view-point.
	
	Now, leveraging on the inferences from our previous work \cite{droidol}, we refrain from using only a subset of these features (e.g., using only the 15,171 features encounter on Day 1). Instead, we allow the dimensionality of our CW classifier to grow with the number of new features encountered; on the last day (13 Aug'14), for instance, we classify with more than 2.1 million features. Implicitly, samples that were introduced before a feature \textit{i} was first encountered will have value 0 for feature \textit{i}. 
	
	DroidOL experimentally demonstrated that the performance for fixed feature regimen is significantly inferior to the growing feature regimen (see \S VI-C of \cite{droidol}). This reveals that continuous retraining with a \textit{growing feature-set} allows a model to successfully adapt to new features on a sub-day granularity. 
	%This adaptiveness is critical to realize the full benefits of online learning. 
	This training regimens helps our CW classifier stay abreast of changing trends in malware and benign apps’ features.
	
	\textbf{Results \& Discussions.}
	Figs. \ref{fig:why-ol} (a), (b) and (c) shows the cumulative error rates of \tool{}  in comparison to the aforementioned variants of SVM, \textsc{Drebin} and CSBD. The following observations are made from the figure:
	\begin{itemize}[leftmargin=*]
		\setlength\itemsep{0em}
		\item Updating the detection models over time is essential to detect new malware as shown by SVM-MultiDaily and SVM-Daily outperforming SVM-MultiOnce and SVM-Once, respectively. 
		
		\item Training on significantly more data improves the performance, as illustrated by SVM-MultiDaily  and SVM-MultiOnce outperforming SVM-Daily and SVM-Once, respectively. However, it is noted that there is a fundamental limit on the amount of data a batch-learning technique could train on because of the storage, memory and time requirements. Thus, among the SVM variants we have considered, SVM-MultiDaily achieves the best accuracy, as it has both the advantages of being trained frequently and trained with large volumes of data.
		
		\item From Fig. \ref{fig:why-ol} (a), one could see that \tool{} consistently outperforms all the batch-learning SVM variants. In particular, when the SVM models are not retrained, \tool{} outperforms SVM-Once and SVM-MultiOnce by more than 32\% and 25\%, respectively. When the SVM models are retrained on a daily basis, \tool{} still outperforms SVM-Daily and SVM-MultiDaily by more than 16\% and 9\%, respectively. 
		%{\color{red} [seems both CWLK and \textsc{Drebin} when retrained performs worse than CSBD. May need to reduce the error rate a bit!]}. 
		This is because \tool{} is able to adapt to the	changes in the malware characteristics instantaneously as well as retain significant useful information from the past. 
		
		\item In the case of comparison with \textsc{Drebin} and CSBD, the trends in the performance of all four variants of these methods are similar to those of the SVM variants. Hence the observations on frequently updating the models and training with more data, hold. %The error rates of best-performing variants of these methods (i.e., \textsc{Drebin}-MultiDaily and CSBD-MultiDaily) are comparable to that of SVM-MultiDaily.
		
		\item From Fig. \ref{fig:why-ol} (b) and (c), one could see \tool{} consistently outperforming the best performing variants of \soa{} methods. Particularly, it outperforms Drebin-MultiDaily by 10.61\% and CSBD-MultiDaily by 7.89\%. This reaffirms the suitability of online learning solutions over retrained batch learners for practical large-scale malware detection.
		
		\item Finally, it is worth noting that the maximum accuracies obtained by \tool{} and other \soa{} methods on ITW dataset are much lesser than those obtained on BM dataset. This substantiates the finding mentioned in \S \ref{ss:ds} that malware detection on large-scale real-world setting is more challenging than the one with benchmark datasets. %Also, in this juncture, it is important to note that \tool{} achieves more than 7\% improvement over the \soa{} methods on the ITW dataset as opposed to 0.15\% improvement on the BM dataset.
	\end{itemize}

	\subsection{RQ2: Efficiency}
	\label{ss:rq2}
	
	%\begin{figure*}[t]
	%	\centering
	%	\includegraphics[height=5.7cm,width=18cm]{time-eff.pdf}
	%	\caption{\small: Comparing \tool's efficiency against that \soa{} methods in terms of training and testing durations \label {fig:eff}}
	%\end{figure*}

	\begin{table}[t]
		\scriptsize
		\centering
		\caption{Comparing \tool's efficiency against that \soa{} methods in terms of training and testing durations}
		\label{tab:eff}
		\begin{tabular}{|l|l|l|}
			\hline
			\textbf{Method} & \textbf{\begin{tabular}[c]{@{}l@{}}Training Duration \\ (avg. $\pm$ std.) in sec.\end{tabular}} & \textbf{\begin{tabular}[c]{@{}l@{}}Testing Duration\\ (avg. $\pm$ std.) in sec.\end{tabular}} \\ \hline
			
			DrebinOnce & \textbf{0.0096} & \textbf{0.0004} ($\pm$ 0.00) \\ %\hline
			DrebinMultiOnce & 0.0352 & 0.0006 ($\pm$ 0.00) \\ %\hline
			DrebinDaily & 0.4493 ($\pm$ 0.08) & 0.0010 ($\pm$ 0.00) \\ %\hline
			DrebinMultiDaily & 0.5873 ($\pm$ 0.27) & 0.0010 ($\pm$ 0.00) \\ \hline
			CSBDOnce & 0.1849 & 0.0354 ($\pm$ 0.01) \\ %\hline
			CSBDMultiOnce & 0.5858 & 0.0594 ($\pm$ 0.03) \\ %\hline
			CSBDDaily & 11.0698 ($\pm$ 5.38) & 0.0605 ($\pm$ 0.02) \\ %\hline
			CSBDMultiDaily & 14.5157 ($\pm$ 11.40) & 0.0641 ($\pm$ 0.03) \\ \hline
			\tool & 0.0131 & 0.0011 ($\pm$ 0.00) \\ \hline
		\end{tabular}
	\end{table}
	
	In this RQ, we investigate the efficiency of \tool{} in terms of training and testing durations. Specifically, we study whether \tool{} achieves superior accuracy at the cost of very high (practically inviable) training or test durations.
	%This study is of paramount importance as \tool{} is posed as a practical, large-scale malware detection framework. 
	
	\textbf{Dataset \& Experiment.}
	The ITW dataset which involves several thousands apps, instigating a vocabulary of over 2 million features for \tool, 1 million features for \textsc{Drebin}, offers enough opportunities to challenge these techniques in terms of efficiency. Hence we consider ITW dataset for this RQ.
	The experimental setting mentioned in RQ1.2 (\S \ref{ss:why-ol}) are reused in this RQ as well.
	
	%In a typical batch learning setting, the classification model is built off-line and is used for predicting the class of the test set samples. 
	A typical batch learning test cycle will involve only feature extraction, representation and prediction on each test set sample. However, in the online learning setting the model will predict and at the same time learn from samples that stream-in. 
	The initial online model is often built with a batch of trivial number of samples and keeps updating itself from the samples that stream-in. 
	%In the case of online learning, the initial model is often built with a batch of trivial number of samples and keeps updating itself from the samples that stream-in. 
	We emulate this scenario, as we use the batch of samples that stream-in on the first day of ITW dataset (i.e., 1 Jan'14) to build the model. The samples that arrive thereafter are considered as stream of samples on which \tool{} predicts the label and also learns from. 
	
	%\textbf{Techniques compared.} The efficiency of all the four variants of \textsc{Drebin} and CSBD discussed in section \S \ref{ss:why-ol}  are compared against those of \tool. 
	
	\textbf{Results \& Discussions.}
	The training and the testing durations of \tool{} and the 4 variants of \soa{} solutions across all the 224 evaluation days in the ITW dataset are presented in Table \ref{tab:eff}.
	% (a) and (b),  respectively.
	In the case of batch learning models, the training durations of variants that are retrained every day (i.e., Daily and MultiDaily) are distributions of values (one for each day). For the variants that are not retrained (i.e., Once and MultiOnce) the training durations are single values, as the training happens only once. 
	%These values are presented in column 2 of Table \ref{tab:eff}. 
	In the case of testing durations, each model undergoes testing cycle every day and hence this produces a series of testing durations. 
	%These testing durations of all the models are furnished in column 3 of the table.
	
	The following observations are made from Table \ref{tab:eff}:
	\begin{itemize}[leftmargin=*]
		\setlength\itemsep{0em}
		
		\item Understandably, the batch learning variants that are trained with smaller sets consume lesser training duration than their counterparts trained with larger sets. For instance, the training durations of Drebin-Once and CSBD-Once are more than 3 times shorter than those of their respective MultiOnce counterparts. Similar trend is observed when we compare the variants that are retrained. That is, on average, both Drebin-Daily and CSBD-Daily are trained 1.3 times faster than their respective MultiDaily counterparts. This shows that the training duration of the batch learning models increase exponentially with the training set size. For obvious reasons, the testing durations of Once and Daily variants are similar to those of their Multi counterparts for both techniques.
		
		\item It is well-known the choice of the classifier plays a pivotal in determining the training and testing durations of the models. Linear models like Linear SVM and CW are simpler and could be trained much faster than quasi-linear models like RFs. This fact is evident from our findings in the aforementioned figure. All the variants of \textsc{Drebin} and \tool{} (which use linear models) are significantly faster than their CSBD counterparts (which use RF classifiers) in terms of both training and testing durations.
		
		\item Finally, we intend to compare \tool's efficiency with those of best performing variants of the \soa{} methods (i.e., Drebin-MultiDaily and CSDB-MultiDaily). The training duration of \tool{} is 44 and 1108 times lesser than those of \textsc{Drebin} and CSBD MultiDaily variants, respectively. \textit{This clearly illustrates that retraining the models at specific intervals are impractical and much less efficient than online learning in the real-world setting}.
		In terms of testing durations, \tool{} is nearly 58 times faster than CSBD-MultiDaily and as fast as Drebin-MultiDaily. 
		%It is noted that unlike Drebin and CSBD, \tool's testing process involves not only predicting the label of the given sample, but also updating the model whenever the predictions are incorrect (see \S \ref{ss:ol} and eq. (5)). We believe this is the main reason for \tool{} consuming more time than Drebin-MultiDaily. However, this update process is so crucial to \tool's effectiveness and makes it nearly 11\% more accurate than Drebin-MultiDaily.
		In summary, on the ITW dataset, \tool{} takes 28.23 microseconds on average to predict the class of a given sample.
	\end{itemize}

	\subsection{RQ3: Explainablity}
	\label{ss:rq3}
	
	\begin{figure}
		\centering
		\includegraphics[height=18cm,width=9cm]{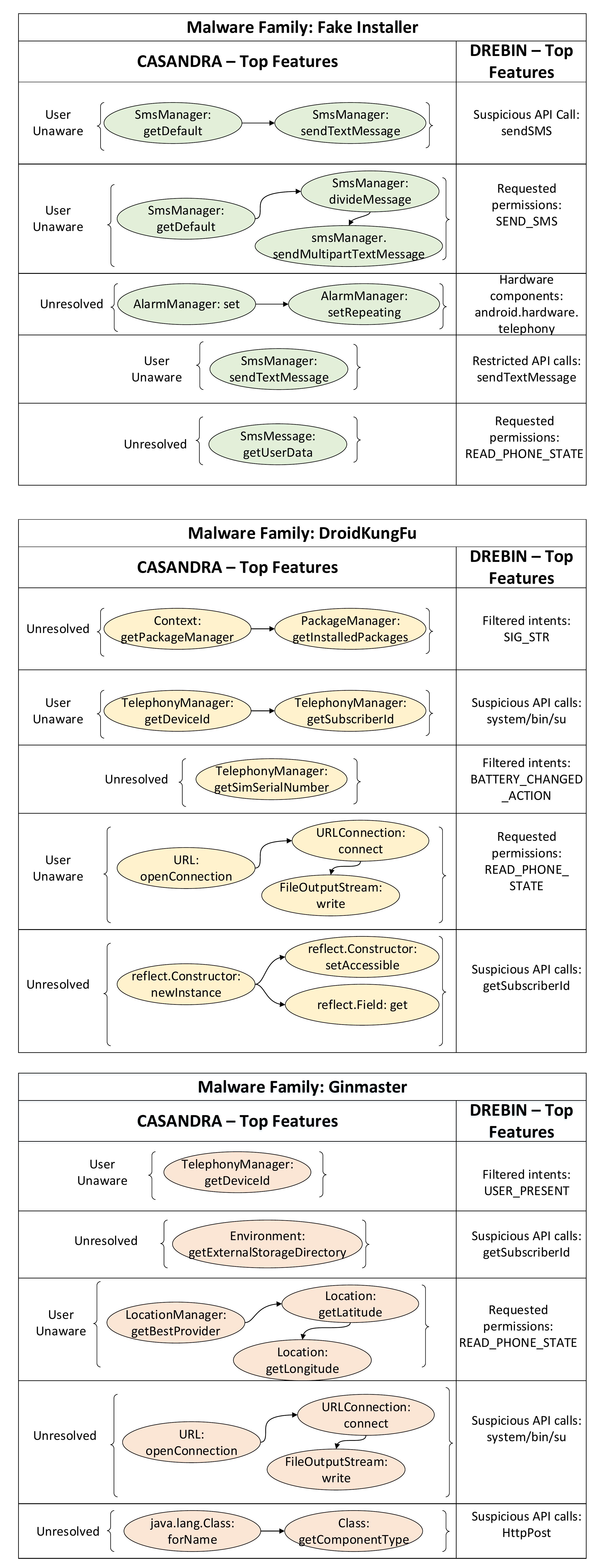}
		\caption{\tool{} Vs. \textsc{Drebin}: Comparing explainability on three popular families from the BM dataset
			\label {fig:exp}}
	\end{figure}
	
	%A typical problem of using machine learning based malware detection approaches is their inability to provide interpretable results. It is often difficult for analysts to understand why samples are marked as benign or malicious. 
	Apart from its detection performance the strength of \tool{} lies in its ability to offer interpretations of the obtained results.
	We verify this quality of \tool{} in this RQ.
	%\tool{} provides insights on the detection by ranking the most important features that correspond to the malicious behavior of an app, making it a malware. 
	%This allows us to check whether these significant features which contribute to the detection correspond to common malware characteristics. 
	%In this section we first take a look at four popular malware families and analyze how features with high weights allow conclusions to be drawn about their behavior. We then inspect false positives and false negatives of MKLDroid in comparison with the state-of-the-techniques in detail.
	
	\textbf{Dataset \& Experiment.}
	In this experiment, we intend to rank the features from a given malware sample that maximally influence \tool's classification decision and investigate if they reflect the malicious behavior of the sample. 
	If they do so, we can conclude that those features explain the sample's behavior the best. 
	Understandably, the ground truth on the malware behavior could be ascertained by knowing the family to which it belongs. For instance, a sample belonging to the \textit{FakeInstaller} family is expected to send premium-rated SMS as this behavior is a part of its attack vector. Therefore, clearly, the availability of malware family labels is indispensable for this experiment. Out of our two datasets, only the BM dataset contains malware family labels. Hence in this RQ, we experiment with three popular malware families from the BM dataset and analyze how \tool{} features with high weights allow conclusions to be drawn about their behavior. 
	
	%\textbf {Techniques compared.}
	As in the previous RQs, we intend to compare the explainability of \tool{} with that of \textsc{Drebin} and CSBD. In terms of classifiers used, it is noted that results of \textsc{Drebin} (which uses linear SVM) and CSBD (which uses RF) are perfectly interpretable. 
	However, the features used by each of these methods are also different and have different levels of interpretability.
	As mentioned in the original work \cite{Drebin}, a significant feature reported by \textsc{Drebin} for the \textit{FakeInstaller} family is the use of permission {\tt SEND\_SMS}. This is highly correlated to the main functionality of the malware. However, for the same family, CSBD, as it considers CFG signatures as features, reports the signature {\tt F0F0F0F0F0F0F0F0F0F0F0P1G} as the most significant feature (see \S 4.1 of \cite{CSBD} for details on extracting CFG signature features). This feature is neither humanly interpretable nor correlated to \textit{FakeInstaller's} malicious  functionality. This clearly illustrates that CSBD's results are inexplicable. Hence, we refrain from including CSBD in this RQ's comparative analysis. 
	
	\textbf{Results \& Discussion.} 
	To study the most significant features that influence the predictions of the \tool{} and \textsc{Drebin}, we consider three well-known malware families, namely \textit{FakeInstaller}, \textit{GingerMaster} and \textit{DroidKungFu} from the BM dataset. For each sample of these families we determine the features with the highest contribution to the classification decision and average the results over all members of a family. The resulting top five features for each malware family for both the techniques are shown in Fig. \ref{fig:exp}. 
	
	From the figures the following inferences are drawn for each family:
	%\begin{itemize}[leftmargin=*]
	%	\setlength\itemsep{0em}
	%\item 
	
	\textit{FakeInstaller} malware samples send premium-rated SMS to specific numbers without users' consent. 
	%These malware hide their attacks inside repackaged versions of popular applications.
	It is evident that features identified by \tool{} not only highlight neighborhoods with operations related to SMS, but also reflect that they are triggered predominantly in the \textit{user-unaware} or \textit{unresolved} context. These operations include getting the default SMS manager, sending normal and multi-part SMS. Also, from the third top feature, one could see that \textit{FakeInstaller} triggers these operations using {\tt AlarmManager} APIs which are invoked in the background without the users' knowledge. On the other hand, \textsc{Drebin}'s explanations are simpler and naive. For instance, as \textsc{Drebin} claims simply sending SMS does not make an app malicious. However, doing the same without the users' consent does. This distinction is not evident from \textsc{Drebin}'s explanations.  
	
	\textit{DroidKungFu} is a sophisticated command-and-control (C\&C) based family of malware capable of exploiting several vulnerabilities in earlier version of Android to gain root access and steal sensitive data from the device. Its intention to leak a variety of private information such as unique identifiers (e.g., IMEI, IMSI) and contents from content provider (e.g., contacts) to the C\&C server over internet is adequately revealed by \tool{}'s features. Also, this malware reads system state changes such as {\tt SIG\_STR} through respective intents and use them to trigger  malicious services in the background. 
	%This behavior is highlighted by the fact that many of \tool's top-ranking malicious subgraphs are reached in the \textit{user-unaware} context. 
	Again, compared to \textsc{Drebin}, \tool's explanations are closer and more descriptive of the secretive operations of \textit{DroidKungFu}. Interestingly, the fact that \textit{DroidKungFu} leverages on \textit{Java reflection} to obfuscate its attack is evident from \tool's explanations. 
	
	\textit{Ginmaster} is a popular Trojan family that involves in leaking users' private information to remote C\&C servers, besides attempting to gain root access.
	Similar to \textit{DroidKungFu}, \textit{Ginmaster} performs its leaks through services that are triggered by background events such as {\tt BOOT\_COMPLETED}. Also, these malware use reflection (through {\tt java.lang.Class} APIs) to obfuscate their attacks. 
	%In summary, their attack and evasion strategies are similar to \textit{DroidKungFu}. 
	Again, a significant part of this behavior can be better reconstructed just by looking at the top features of \tool{}.
	
	%\end{itemize}

	\begin{figure*}[t]
		\centering
		\includegraphics[height=10cm,width=18cm]{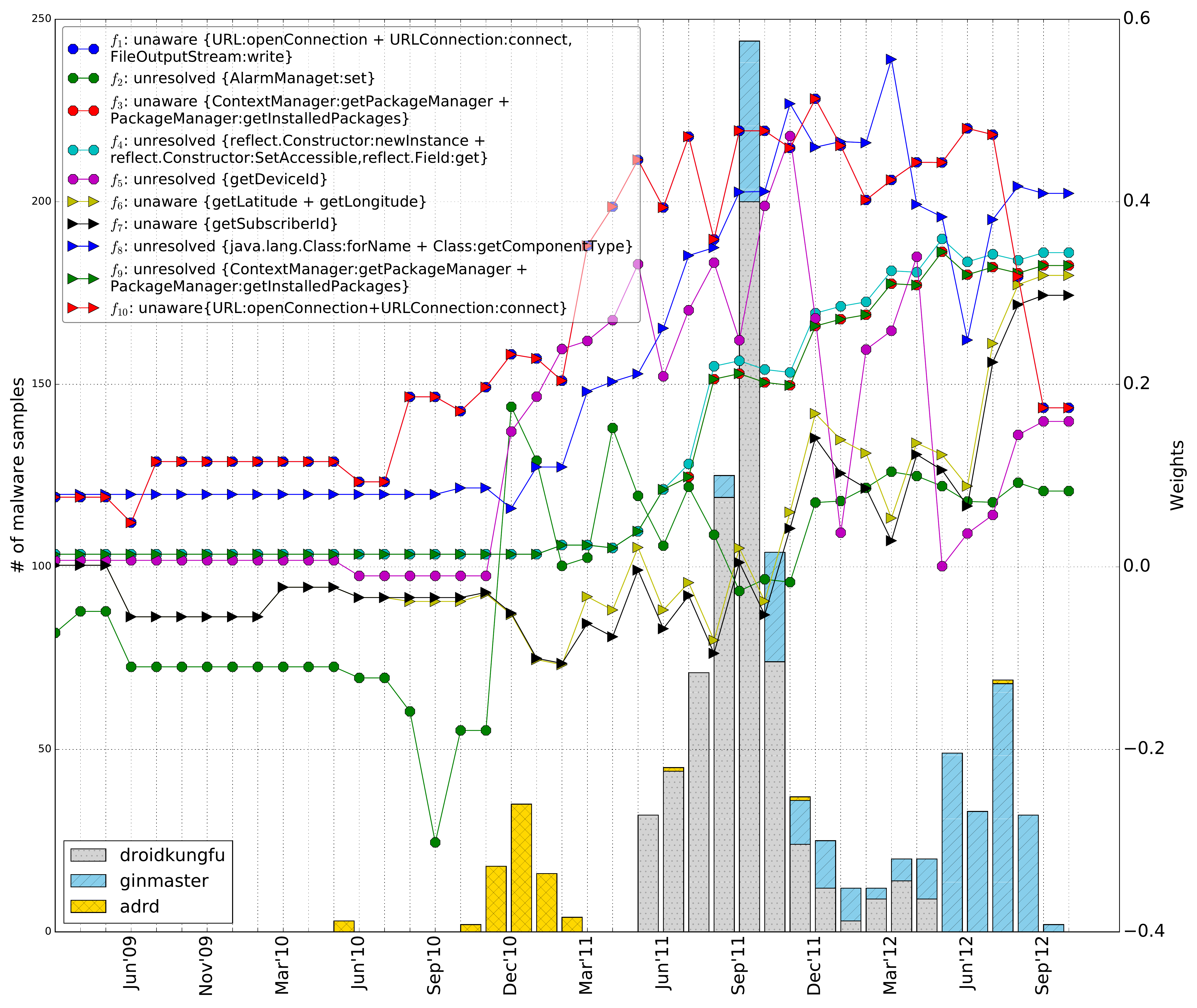}
		\caption{\small Illustration of how \tool{} adapts to the malware evolution and population drift in the BM dataset. The X axis shows the distribution of family-specific malware samples. Y axis (left) shows the number of samples and Y axis (right) shows the weights of individual features. The legend contains CWLK subgraph features (format: $ < $feature id$ > $: $ < $context$ > $ \{$ < $root-node$ > $ + $ < $neighbor-list$ > $\})}
		\label{fig:ta-1}
	\end{figure*}
	
	\subsection{RQ4: Adaptiveness}
	\label{ss:rq4}
	%\begin{figure}
	%	\centering
	%	\includegraphics[height=5.5cm,width=10cm]{40PercentMonthlyPorportion.png}
	%	\caption{Date of creation of \textsc{Drebin} malware samples according to family
	%		\label {fig:evol_in_drebin}}
	%\end{figure}

	%\begin{figure*}[t]
	%	\centering
	%	\includegraphics[height=12cm,width=18cm]{SMS.pdf}
	%	\caption{\small: Trend Analysis on premium-rated SMS family \label {fig:ta-2}}
	%\end{figure*}

	In this RQ, we intend to study how \tool{} adapts itself to malware evolution and population drift. Specifically, we intend to explore how online learning helps \tool{} to learn fresh patterns and unlearn obsolete patterns of malicious behavior over time. As stated earlier, all the existing Android malware detection approaches (incl. \textsc{Drebin} \cite{Drebin} and CSBD \cite{CSBD}) are based on batch learning and are not capable of adapting themselves to population drift. Adaptiveness is \tool’s unique feature and hence we could not compare this with any existing technique. Hence, we just illustrate how \tool{} achieves adaptiveness in this subsection.

	\textbf{Dataset \& Experiment.} In this RQ we prefer the BM dataset for three reasons: (1) it hosts malware collection over a longer period which is ideal to study malware evolution (2) its heterogeneity – it contains malware with attacks as simple as sending premium-rated SMS to complex botnets, (3) availability of accurate malware family labels.
	% which supports inferring the time-line in which families emerge and vanish. 
	%The family labels are not available for the malware in the ITW dataset.
	
	In this experiment, the apps in BM dataset are sorted according to their month of creation. We observe that several malware families in this dataset emerge, flourish and fade-away over time due to various domain-specific reasons. For example, \textit{DroidKungFu} family first emerged in May’11 and reached its peak spread during Sep-Oct’11 and gradually disappeared by May’12. %Similarly, \textit{FakeInstaller}, an equally populous family, but different from \textit{DroidKungFu} in terms of attacks emerged in Sep’11 and existed till the last month of \textsc{Drebin}'s collection. 
	
	Any good malware detector must adapt to such evolution. % and at the same time retain significance of features that are relevant for detection over a long time. 
	To  examine whether \tool{} exhibits this adaptiveness, we train it in an online fashion on the BM dataset following a procedure similar to RQ1.2 (\S \ref{ss:why-ol}). The training and prediction quantum changes from days in RQ1.2 to months in this RQ. 
	
	Once the prediction and learning for all the samples in a month $m$ is over, we record the weights of all the features. Since BM dataset hosts malware collection over 44 months, the weights of each feature are recorded at 44 points in time. This gives us an opportunity to monitor and track the fluctuations in the importances of individual features over time. In general, features that characterize typical malicious and benign behaviors will be assigned large positive and negative weights, respectively. Features that do not reveal anything about malice or benignity will be assigned near-zero weights.
	
	For a detector that adapts well to the evolving trend in malware population, the feature weights should follow the pattern of population drift. More specifically, when a particular family of malware $\mathcal{F}$ gains prominence over a period, say from month $m$ to $m'$, the weights of features that characterize the attacks and evasion strategies of $\mathcal{F}$ should  increase or remain high. Once the samples from family $\mathcal{F}$ start to dwindle, the weights of those features should decrease or remain low. 
	
	\textbf{Results \& Discussion.} 
	In order to study this behavior, we chose three malware families that perform privacy leak attacks, namely, \textit{DroidKungFu}, \textit{Ginmaster}, and \textit{ADRD}. The behaviors of \textit{DroidKungFu} and \textit{Ginmaster} are explained in the previous RQ. \textit{ADRD} also secretively leaks users' private data in a similar fashion. As shown in Fig. \ref{fig:ta-1}, in the BM dataset, \textit{ADRD} first emerged in May'10 and gained prominence from Oct'10 to Dec'10. It finally vanished from Mar'11. A very small number of \textit{ADRD} samples resurfaced in Jun'11, Nov'11, and Jul'12. The evolution of \textit{DroidKungFu} and \textit{Ginmaster} populations could be inferred in a similar fashion from Fig. \ref{fig:ta-1}. 
	%\textit{DroidKungFu} emerged in May'11, gained rapid prominence till Sep'11 and finally vanished after Apr'12. Similarly, \textit{Ginmaster} emerged in Aug'11 and gained prominence in two cycles (Sep-Oct'11 and May-Jul'12) and finally vanished by Sep'12. 
	Now, leveraging on \tool's explainability, we choose 10 features that best explain the privacy leaks from these families and study fluctuations in their weights over time. These fluctuations are also shown in Fig. \ref{fig:ta-1}.
	
	One could clearly see that the feature weight fluctuations follow the evolution of the corresponding families. For instance, from Mar'09 to Apr'10 many of these features had no significant positive weights as none of these populous privacy leak families emerged. Some features 
	%(e.g., {$f_7$}) 
	had near-zero weights indicating that they do not characterize any malice at that point in time. 
	%The only set of exceptions is the URL related features. They possess moderate positive weights as they are common features across many malware families. 
	After May'10, a small number of features that characterize \textit{ADRD} start to gain positive weights. After the voluminous influx of \textit{DroidKungFu} in May'11, almost all the features started to gain larger positive weights. 
	In particular, one could see the following interesting patterns:
	\begin{itemize}[leftmargin=*]
		\setlength\itemsep{0em}  
		
		\item Feature $f_2$ typically characterizes the \textit{ADRD} family. The weights of this feature surge rapidly during \textit{ADRD}'s peak spread period (i.e.,Oct-Dec'10).
		
		\item Features related to reflection (i.e., $f_4$ and $f_8$) and inferring installed apps (i.e., $f_3$ and $f_9$) are common for both \textit{DroidKungFu} and \textit{Ginmaster}. These features receive significant increase in their weights once \textit{DroidKungFu} emerges in Jun'11. Even after \textit{DroidKungFu} vanishes, these features continue to accumulate more weights as \textit{Ginmaster} samples keep streaming till Sep'12.
		
		\item The weights of features that characterize common malicious behaviors across many families such as URL features (i.e., $f_1$ and $f_{10}$) remain predominantly high throughout the entire duration.
		
		\item Similar subgraph features exhibit very close or exactly same weight fluctuation patterns. For instance, the two {\tt PackageManager} features (i.e.,$f_3$ and $f_9$) and two URL features (i.e., $f_1$ and $f_{10}$) have exactly same pattern over the entire duration. This reflects the fact that these features characterize either same or very similar attacks. 
		
		%\item Overall, the weights of all these 10 features keep predominantly increasing over time. Many of them start to increase and reach positive weights when \textit{ADRD} and \textit{DroidKungFu} flourish. Subsequently, after \textit{Ginmaster} arrives in Aug'11, the weights of all these features remain entirely positive.
		
		\item Finally, when we reach the end of the dataset collection period in Sep'12, we could see \tool{} has assigned large positive weights to all the 10 features. Meaning, it has learned that all of them characterize malicious behaviors. In particular, \tool{} has learnt that features such as {\tt reflection} and {\tt PackageManager} APIs ($f_3$ and $f_9$) characterize stronger malicious behaviors than other features such as reading device ID ($f_5$), which may be common in benign apps as well. This is reflected from the fact that it has assigned larger positive weights to the former set of features than the latter.
	\end{itemize}
	
	%In particular, important features related to reflection (e.g., {\tt reflection.Constructor} and {\tt java.lang.Class: forName} related features) and accessing information on installed apps (e.g., {\tt getInstalledPackages} related features) that typically characterize \textit{DroidKungFu} gain significant positive weights in this small interval.  
	
	In summary, we could clearly see that \tool{} when trained in an online fashion, adapted well to the population drift and evolution in malware. So far, none of the existing techniques (incl. \textsc{Drebin} and CSBD) demonstrated this capability. We believe adaptiveness is the cornerstone for the practical success of ML based malware detector.
	
	\section{Related Work}
	\label{sec:rw}
	
	\subsection{Android Malware Detection}
	\label{ss:amt}
	
	Many well-known malware detection solutions have been reviewed in the previous sections. We throw light on remaining works and contrast them from \tool{} in this subsection.
	%We review well-known ML based malware detection solutions in this subsection. 
	Features used for detection is the most important aspect of ML based malware detectors. Hence, we discuss the related work with respect to this aspect. 
	
	%The existing solutions predominantly rely on either static analysis or dynamic analysis for modeling the behaviors of apps, subsequently extract semantic features and finally use ML classifiers train and predict malware. 
	%Since we have extensively explained \textsc{Drebin} \cite{Drebin} and CSBD \cite{CSBD} in \S \ref{ss:impl}, we exclude them from further discussions here.
	
	Crowdroid \cite{Crowd} leverages on dynamic analysis and uses Linux system-call sequences as features. 
	DroidAPIMiner \cite{droidapiminer} considers a hand-crafted set of sensitive APIs (along with their parameters) and package level information as features. 
	MAST \cite{mast} uses selected permissions, Intent filters, the existence of native code and zip files, then applies Multiple Correspondence Analysis to perform malware detection. 
	Peiravian et al. \cite{peir}  take a simpler approach by considering permissions and API calls as features. 
	\textsc{Chabada} \cite{chabada} leverages on text-mining techniques to study the relevance between apps’ behavior and their description and henceforth detect suspiciously behaving apps. 
	Sahs and Khan \cite{MLMalDetect} extract a variety of features including permissions and CFG signatures and perform early-fusion of those features. With these features, they take an anomaly detection approach using a One-Class Support Vector Machine to detect malware.
	\textsc{Adagio} \cite{Adagio} constructs apps' call-graphs (CGs) and uses byte-code instructions to assign labels to nodes. Subsequently, it captures structural information from CGs using NHGK and uses kernel SVM to detect malice.
	DroidMiner \cite{DroidMiner} proposes a two-tiered behavior graph to model malicious program logic into a vector of threat modalities, and then applies classification according to these modalities. 
	%DroidSIFT \cite{DroidSift} further models API-relevant behaviors into weighted contextual API dependency graphs and classifies malware based on these graphs. However, unlike \tool{}, DroidSIFT uses a fixed vocabulary of hand-picked subgraphs as the vocabulary to construct feature vector representations.
	%Recently, AppContext  proposes differentiating malicious and benign behaviors based on the contexts similar to ours. However, it ends up capturing contextual features from individual nodes without their topological structural information.
	MassVet statically analyzes apps' UI code to extract a graph that expresses the UI state and transitions. Subsequently, it uses a centroid-based approach to detect repackaged apps and DiffCom analysis to identify the malicious portions of those apps.
	
	Recently, several techniques such as MARVIN \cite{marvin}, HADM \cite{hadm} and StormDroid \cite{stormdroid} adopt a hybrid approach by combining both static and dynamic analysis features and demonstrated achieving better accuracies than solutions leveraging on one of the analysis paradigms.
	
	However, all these solutions are based on batch-learning and they neither account for concept drift in Android malware nor exhibit adaptiveness. On the contrary, \tool{} uses online learning and is naturally adaptive to concept drift. Moreover, none of the above-mentioned techniques exhibit all the four qualities that \tool{} possess (i.e., accurate context-aware detection, efficiency, explainability and adaptiveness). The only other work that discusses concept drift in Android malware is Prescience \cite{prescience}. It detects drifts in malware concept using Venn-Abers predictors and retrains the computationally heavy models such as XGBoot and ExtraTrees periodically. Though, this retraining approach addresses concept drift, as we showed in \S \ref{ss:rq2} it is significantly less efficient than using online learning.
	
	\subsection{Graph Kernels}
	\label{ss:gk}
	Several general purpose graph kernels have been proposed and used in many real-world applications. 
	These graphs kernels can be categorized into four major families: \\
	\noindent
	\textbf{1. Graph kernels based on walks/paths.}
	Random Walk (RW) \cite {rw} and Shortest Path (SP) \cite{sp} graph kernels are classic examples of this family. The RW kernel counts the number of common walks when simultaneous walks are performed on a pair of graphs. The SP kernel simply compares the sorted endpoints and the length of shortest-paths that are common between two graphs. A significant bottleneck for both these kernels is their computational cost. The best known time complexity for exact computation is $O(n^3)$ for both RW and SP kernels.\\ %Attempts have been made to reduce the runtime. For instance Kang et al. \cite{frw} introduce low-rank approximation based computation and achieve better complexity.
	\noindent
	\textbf{2. Graph kernels based on subgraphs.}
	The graphlet \cite{graphlet} and Neighborhood Subgraph Pairwise Distance kernel (NSPDK) \cite{NSPDK} are popular instances of this family.
	Graphlet kernel counts the common non-isomorphic subgraphs (upto a certain degree) among a pair of graphs to evaluate their similarity and is computable in polynomial time.
	NSPDK counts the number of rooted subgraphs containing nodes up to a certain distance from the root, the roots of which are located at a certain distance from each other, in two graphs.\\
	\noindent
	\textbf{3. Graph kernels based on subtree patterns. }
	WLK is the most popular instance of this family. Since we have explained it extensively in several sections, we refrain from discussing it here. \\
	\noindent
	\textbf{4. Deep Graph Kernels.}
	Recently, the success of deep representation learning techniques in other fields like %Computer Vision and 
	Natural Language Processing inspire researchers to learn representations of sub-structures from graph and thus develop deep learning variants of popular graph kernels.
	For instance, Yanardag and Vishwanathan \cite{dgk} develop deep learning variants of many of the above-mentioned kernels such as graphlet, SP and WLK. Similarly, Narayanan et al. \cite{sg2vec}  learn latent representations of rooted subgraphs and thus develop a deep learning variant of WLK.
	
	Considering the malware detection application, while all these general purpose kernels could capture and compare structural data well, they fail to capture reachability contexts as it is a strong application-specific requirement. CWLK is specifically designed to address this research gap.\\
	\noindent
	\textbf{Application-specific graph kernels.}
	Like CWLK, several application-specific kernels have been proposed in the past. 
	%De Vries et al \cite{} propose an extension to WLK for handling Resource Description Framework data and cater relevant applications such as affiliation and theme prediction.
	For instance, Harchaoui and Bach \cite{CVGK} propose a  segmentation kernel that counts common virtual sub-structures amongst images to perform Computer Vision tasks such as image classification.
	Mahe et al. \cite{Molec} propose a marginalized graph kernel for a chemoinformatics application where they enrich the node labels using the Morgan index and counted the number of matching nodes across graphs. Driven by similar motivation, we have developed CWLK to cater specifically to malware detection. To the best of our
	knowledge, this is the first graph kernel specifically addressing
	a problem from the field of program analysis.
	
	%\subsection{Online Learning}
	%\label{ss:rw_ol}
	
	\section{Limitations}
	\label {sec:lim}
	\noindent
	\textbf{Static Analysis Limitations.} The previous evaluation demonstrates the efficacy of our method in detecting recent malware on the Android platform. However, \tool, cannot generally detect all sorts of malicious behaviors, as it builds on concepts of static analysis and lacks dynamic inspection. In particular, transformation attacks that are non-detectable by static analysis (e.g., attacks based on reflection and bytecode encryption could go undetected). To alleviate the absence of a dynamic analysis, \tool{} extracts API calls related to obfuscation, reflection, and loading of code, such as {\tt reflect.Constructor} and {\tt DexClassLoader.loadClass}. These features enable us to at least spot the execution of hidden code—even if we cannot further analyze it. In combinations with other features, \tool{} is still able to identify malicious behaviors despite the use of some obfuscation techniques.\\ %Also, leveraging on recent works such as DroidRA \cite{droidra} and \cite{droidnative} we intend to address reflection and native code based malware in our future work.\\
	\noindent
	\textbf{Adversarial attacks.} Another limitation which follows from the use of ML is the possibility of attacks by adversaries such as poisoning (see \cite{poison,adverdrift}). While common obfuscation strategies, such as identifier renaming and code reordering do not affect \tool, adversaries may succeed in reducing its accuracy by incorporating benign CADG subgraph features or fake invariants into malicious apps.
	When attackers succeed in tampering online learning methods and make them adapt to carefully crafted noise, the damages caused are much worse compared to batch learners as they may continuously fit to noise allowing malware to go undetected.
	Even though such adversarial attacks against ML based detectors cannot be ruled out in general, meticulous sanitization of training data (see \cite{adverdrift}) can limit their impact.
	%\noindent
	%\textbf{Availability of labels.} TBD.

	\section{Conclusion \& Future work}
	\label{sec:conc}
	In this paper, we present \tool, an online learning based Android malware detection framework. CWLK, a novel graph kernel that facilitates capturing apps' security-sensitive behaviors along with their context information from dependency graphs is also proposed. 
	CWLK supports explicit feature vector representation of apps' dependency graphs using which an online classifier is trained to detect malware. 
	Our large-scale evaluations on both recent real-world and benchmark datasets demonstrate that \tool{} outperforms two \soa{} techniques. \tool{} achieves 89.92\% accuracy on a real-world dataset with more than 87,000 apps outperforming \soa{} techniques by more than 25\% in their typical batch-learning setting. 
	On average, \tool{} takes 28.23 microseconds to predict the label of a given sample in 
	our large-scale experiment, which is comparable to state-of-the-art techniques, making it scalable enough to perform market-scale analysis. This superior performance and scalability make \tool, in particular, and online learning based solutions, in general, better candidates for the malware detection task. 
	
	\textbf{Future Work.} Taking into account the recent developments in the area of Deep Graph Kernels (e.g., \cite{dgk,sg2vec}), which show potentials to learn latent sub-structures from graphs to achieve better accuracy, we intend to explore on the deep learning variant of CWLK in our future work.
	%and (ii) inspired by recent works such as \cite{hadm,stormdroid} we intend to combine both static and dynamic analyses and build multiple types of dependency graphs in our feature engineering phase. We plan to leverage on online multiple kernel learning towards integrating them. 
	
	\appendices
	
	\section {Primer on Kernel Methods \& Graph Kernels}
	\label{app:kmgk}
	Kernel methods have been highly successful in solving a specific class of problems where feature vector representations of samples are not readily obtainable. Malware detection using PRGs is one of such problems. For many well-known classifiers, the data samples have to be explicitly represented as feature vectors through a user-specified feature map $\phi$. In contrast, kernel methods require only a user-specified kernel $k$, i.e., a similarity function over pairs of samples in their native representations. Kernel methods work by mapping the samples into a feature space, implicitly and finding an appropriate decision boundary in the new feature space. Here, feature map $\phi(\cdot)$ is realized through the kernel function $k$, which facilitates computing inner products in the feature space using the samples in their native representation, i.e. $k(x_i, x_j ) = \langle \phi (x_i), \phi (x_j) \rangle$. 
	%The kernel function must be positive definite for most kernel classifiers. Examples of positive definite kernels are the Dirac, Histogram Intersection and Gaussian kernels.
	%Roughly speaking, \textit{a kernel value is a measure of similarity between a pair of samples}. %Therefore, a graph kernel is a similarity measure between graphs.
	
	\textbf{Graph Kernels.}
	Formally, a graph kernel $k : \mathbb{G} \times \mathbb{G} \rightarrow R$ is a kernel function defined on a domain of graphs, $\mathbb{G}$. Hence, given a labeled graph dataset $D_g = {(g_1, y_1),..., (g_n, y_n)}$ and a graph kernel $k$, a kernel classifier (e.g. SVM) can be directly used to perform graph classification. Graph kernels usually belong to the family of \textit{R convolution kernels}. These kernels decompose graphs into sub-structures such as paths, walks etc. The comparison of two graphs is then based on the similarity between all pairs of such sub-structures. Several graph kernels have been proposed based on this idea \cite{rw,sp,frw,WLK,NHGK,NSPDK}. %, for instance,\textit{ random walks kernels, shortest paths kernels} and Weisfeiler-Lehman (WL) kernels \cite{WLK}.
	
	%Usually, the choice of a graph kernel for a particular task depends on three factors:\\
	%\textbf{Scalability.} Since graphs are complex data structures with node/edge labels and attributes computing similarity among their sub-structures is often not scalable. For instance, classical graph kernels such as random-walk kernels, despite their nice theoretical properties, are computationally expensive making them unsuitable for a problem like malware detection where we process large volumes of graphs with thousands of nodes and edges. On the other hand scalable graph kernels such as WL and LTGK have been successfully used for malware detection at scale in the past.
	
	\textbf{Explicit vs. Implicit Feature mapping.} Existing graph kernels can be classified into approaches that use explicit feature mapping ($\phi$) and those that directly compute a kernel function (i.e., $\phi$ is not necessarily known and may be of infinite dimension). Examples of former category include WLK \cite{WLK} and NHGK \cite{NHGK}, and that of latter category are RW \cite{rw} and SP \cite{sp} kernels. Kernels that support explicit feature mapping have two advantages that make them particularly suitable for the malware detection problem:\\ 
	\textit{(1) Scalability.} If explicit representations are manageable, these approaches usually outperform other kernels regarding runtime on large datasets, since the number of vector representations scales linear with the dataset size.\\ 
	\textit{(2) Explainability.} These kernels support extracting sub-structures of graphs as features and building a vocabulary of such features. This facilitates building explicit feature vector representation of individual graphs.
	% Therefore, both kernel and non-kernel based classifiers could be used for classification. 
	This aspect makes this category of kernels amenable for performing explainable malware detection.
	
	Two explicit feature mapping kernels, namely, WLK \cite{WLK} and NHGK \cite{NHGK} have been successfully used for Android malware detection in \cite{MLMalDetect} and \cite{Adagio} respectively. Moreover, \cite{Adagio} performed explainable detection leveraging on the explicit feature map produced by NHGK.

	\section{CWLK positive semi-definiteness}
	\label{app:mercer}
	\textbf{Theorem 1.} CWLK is positive definite.\\
	\textbf{Proof.} Let us define a mapping $\phi$ that counts the occurrences of a particular contextual neighborhood label sequence $\gamma$ in $G$ (generated in $h$ iterations of Algorithm \ref{algo:cr}). Let $\phi^{(h)}_\gamma(G)$ denote the number of occurrences of $\gamma$ in $G$, and analogously $\phi^{(h)}_\gamma(G')$ for $G'$. Then,
	\begin{equation}
	\begin{aligned}
	k^{(h)}_{\gamma}(G, G') = \phi^{(h)}_\gamma(G), \phi^{(h)}_\gamma(G') = |\{(\gamma_i(n),\gamma_i(n'))|\\ \gamma_i(n) = \gamma_i(n'), i \in \{0,...,h\}, n \in N, n' \in N'\}|
	\end{aligned}
	\end{equation}
	\vspace{-2mm}
	Summing over all $ \gamma $ from the vocabulary $ \Sigma^* $, we get
	{\small \begin{multline}
		k^{(h)}_{CWL}(G, G') = \sum_{\gamma \in \Sigma^*}^{} k^{(h)}_{\gamma}(G, G') = \sum_{\gamma \in \Sigma^*} \phi^{(h)}_\gamma(G)\phi^{(h)}_\gamma(G')\\ = |\{(\gamma_i(n),\gamma_i(n'))|\gamma_i(n) = \gamma_i(n'), i \in \{0,...,h\}, n \in N, n' \in N'\}|
		\end{multline}}
	\nolinebreak
	In general, $ k^{(h)}_{CWL} $ defines a kernel with corresponding feature map $ \phi^{(h)}_{CWL} $, such that
	\begin{equation}
	\phi^{(h)}_{CWL} = (\phi^{(h)}_{\gamma} (G))_{\gamma \in \Sigma^*} %= (\phi_{l}^{(h)}(G))_{l \in }
	\end{equation}
	
	For proof of validity of CWLK with compression on contextual neighborhood labels using a function $f: \Sigma^* \rightarrow \Sigma$ such that $f(\gamma_{i}(n)) = f(\gamma_{i}(n'))$, iff $\gamma_{i}(n) = \gamma_{i}(n')$ could be found in \S IV of \cite{cwlk}.
	
	\section{CWLK Time Complexity}
	\label{app:tc}
	
	The runtime complexity of CWLK with $h$ iterations on a graph with $n$ nodes and $e$ edges is $O(he)$ (assuming $ e\! > \! n $) which is same as that of WLK.
	More specifically, the neighborhood label computation with sorting operations (lines 10-12  of Algorithm \ref{algo:cr}) take $ O(e) $ time for one iteration and the same for $ h $ iterations take $ O(he) $.
	The inclusion of context (lines 6-8,13-15), does not incur additional overhead as $ e\! >\! n$. Hence the final time complexity remains as $O(he)$. For a detailed derivation and analysis of the time complexity of WLK, we refer the reader to \cite{WLK}.
	
	\section {Feature space growth}
	\label{app:fg}
	\begin{figure}[ht]
		\centering
		\includegraphics[height=6cm,width=9cm]{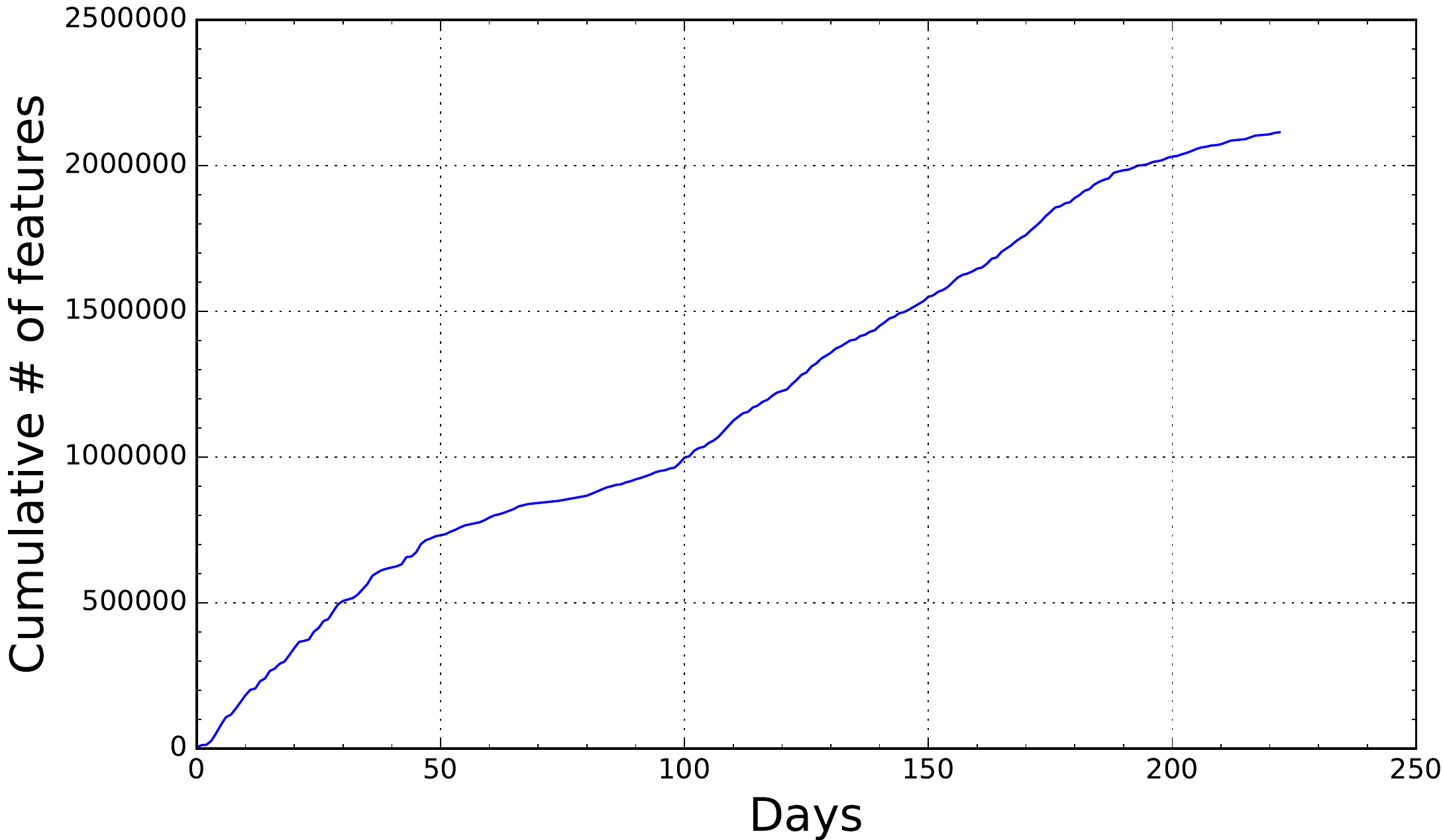}
		\caption{Cumulative number of features observed over time for
			our ITW dataset.
			\label {fig:feat_growth}}
	\end{figure}

\end{document}